\documentclass[dvips]{llncs}

\usepackage{t1enc}
\usepackage[latin2]{inputenc}
\usepackage[english]{babel}
\usepackage{amsmath}
\usepackage[mathscr]{eucal}
\usepackage{graphicx}
\usepackage{subfig}
\usepackage{bm}
\usepackage{url}
\usepackage{cite}
\usepackage{amssymb}

\usepackage[all]{xy}
\usepackage{rotating}
\usepackage{pstricks}
\usepackage{pst-fr3d}
\newcommand{\myar}{\ar@[|(1.7)]}

\newcommand{\R}{\ensuremath{\mathbb{R}}}

\spnewtheorem*{mainthm}{Theorem}{\bfseries}{\itshape}
\renewcommand{\b}{\mathbf}
\title{Undercomplete Blind Subspace Deconvolution via Linear Prediction}
\author{Zolt{\'a}n Szab{\'o} \and Barnab{\'a}s  P{\'o}czos \and Andr{\'a}s L{\H{o}}rincz}
\institute{Department of Information Systems, E\"{o}tv\"{o}s Lor{\'a}nd University,\\
P{\'a}zm{\'a}ny P. s{\'e}t{\'a}ny 1/C, Budapest H-1117, Hungary\\
WWW home page: \url{http://nipg.inf.elte.hu} (\url{http://nipg.info})\\
\email{szzoli@cs.elte.hu, pbarn@cs.elte.hu, andras.lorincz@elte.hu}}

\begin{document}

\maketitle

\begin{abstract}
We present a novel solution technique for the blind subspace
deconvolution (BSSD) problem, where temporal convolution of
multidimensional hidden independent components is observed and the
task is to uncover the hidden components using the observation
only. We carry out this task for the undercomplete case (uBSSD):
we reduce the original uBSSD task via linear prediction to
independent subspace analysis (ISA), which we can solve. As it has
been shown recently, applying temporal concatenation can also
reduce uBSSD to ISA, but the associated ISA problem can easily
become `high dimensional' \cite{szabo07undercomplete}. The new reduction method circumvents
this dimensionality problem. We perform detailed studies on the
efficiency of the proposed technique by means of numerical
simulations. We have found several advantages: our method can
achieve high quality estimations for smaller number of samples and
it can cope with deeper temporal convolutions.
\end{abstract}

\section{Introduction}\label{sec:intro}
There is a growing interest in independent component analysis (ICA) and blind source deconvolution (BSD) for signal
processing and hidden component searches. ICA has been used for many purposes, including (i) feature extraction, (ii)
denoising, (iii) processing of financial and neurobiological data, e.g. fMRI, EEG, and MEG. BSD has also shown
potentials in several areas, for example (i) in remote sensing applications: passive radar/sonar processing, (ii) in
image-deblurring and image restoration, (iii) in acoustics, including speech enhancement using microphone arrays, (iv)
in multi-antenna wireless communications and in sensor networks, (v) in biomedical signal---EEG, ECG, MEG,
fMRI---analysis, (vi) in optics, and (vii) in seismic exploration. For recent reviews in ICA and BSD themes see, e.g.,
\cite{hyvarinen01independent,cichocki02adaptive} and \cite{pedersen07survey}, respectively.

Traditionally, ICA is one-dimensional in the sense that all sources are assumed to be independent real valued
stochastic variables. The traditional example of ICA is the so-called \emph{cocktail-party problem}, where there
are $D$ sound sources and $D$ microphones and the task is to separate the original sources from the observed
mixed signals. Clearly, applications where not all, but only certain groups of the sources are independent may
have high relevance in practice. In this case, independent sources can be multidimensional. For example, there
could be \emph{independent groups of people} talking about independent topics at a conference, or
\emph{independent rock bands} may be playing at a party. This is the independent subspace analysis (ISA)
extension of ICA \cite{cardoso98multidimensional}. Strenuous efforts have been made to develop ISA algorithms,
where the theoretical problems concern mostly (i) the estimation of the entropy or of the mutual information, or
(ii) joint block diagonalization. A recent list of possible ISA solution techniques can be
found in \cite{szabo07undercomplete}.

Another extension of the original ICA task is the BSD problem \cite{pedersen07survey}, where the observation is
a temporal mixture of the hidden components. Such a problem emerges, e.g., if the cocktail-party is held in an
\emph{echoic room}. A novel task, the blind subspace deconvolution (BSSD) \cite{szabo07undercomplete} arises if
we combine the ISA and the BSD assumptions. One can think of this task as the separation problem of the pieces
played simultaneously by independent rock bands in an echoic stadium. One of the most stringent applications of
BSSD could be the analysis of EEG or fMRI signals. The ICA assumptions could be highly problematic here, because
some sources may depend on each other, so an ISA model seems better. Furthermore, the passing of information
from one area to another and the related delayed and transformed activities may be modeled as echoes. Thus, one
can argue that BSSD may fit this important problem domain better than ICA or even ISA. It has been shown in
\cite{szabo07undercomplete} that the undercomplete BSSD task (uBSSD)---where in terms of the cocktail-party
problem there are more microphones than acoustic sources---can be reduced to ISA by means of temporal
concatenation.\footnote{The complete, and in particular the overcomplete BSSD task is challenging and no general
solution is known yet.} However, the reduction technique may lead to `high dimensions' in the associated ISA
problem. Here, an alternative reduction method solution is introduced for uBSSD and this solution avoids the
increase of ISA dimensions. Namely, we show that one can apply the linear prediction method to reduce the uBSSD
task to ISA such that the dimension of the associated ISA problem equals to the dimension of the original hidden
sources. As an additional advantage, we shall see that this reduction principle is more efficient on
problems with deeper temporal convolutions.

The paper is built as follows: Section~\ref{sec:BSSD-model}
formulates the problem domain. Section~\ref{sec:reduction-steps}
shows how to reduce the uBSSD task to an ISA problem.
Section~\ref{sec:illustrations} contains the numerical
illustrations. Section~\ref{sec:conclusions} contains a short
summary.

\section{The BSSD Model}\label{sec:BSSD-model}
We define the BSSD task in Section~\ref{sec:BSSD-Eqs}. Earlier BSSD reduction principles are reviewed in
Section~\ref{sec:BSSD-red-principles}.

\subsection{The BSSD Equations}\label{sec:BSSD-Eqs}
Here, we define the BSSD task. Assume that we have $M$ hidden,
independent, multidimensional \emph{components} (random
variables). Suppose also that only their casual FIR filtered
mixture is available for observation:
\begin{equation}
\b{x}(t)=\sum_{l=0}^L\b{H}_l\b{s}(t-l),\label{eq:obs-1}\\
\end{equation}
where
$\b{s}(t)=\left[\b{s}^1(t);\ldots;\b{s}^M(t)\right]\in\R^{Md}$ is
a vector concatenated of components $\b{s}^m(t)\in\R^{d}$. Here,
for the sake of notational simplicity we used identical dimension
for each component. For a given $m$, $\b{s}^m(t)$ is i.i.d.
(independent and identically distributed) in time $t$, there is at
most one Gaussian in $\b{s}^m$s, and
$I(\b{s}^1,\ldots,\b{s}^M)=0$, where $I$ stands for the mutual
information of the arguments. The total dimension of the
components is \mbox{$D_s:=Md$}, the dimension of the observation
$\b{x}$ is $D_x$. Matrices $\b{H}_l\in \R^{D_x\times D_s}$
$(l=0,\ldots,L)$ describe the convolutive mixing. Without any loss
of generality it may be assumed that $E[\b{s}]=\b{0}$, where $E$
denotes the expectation value. Then $E[\b{x}]=\b{0}$ holds, as
well. The goal of the BSSD problem is to estimate the original
source $\b{s}(t)$ by using observations $\b{x}(t)$ only. The case
$L=0$ corresponds to the ISA task, and if $d=1$ also holds then
the ICA task is recovered. In the BSD task $d=1$ and $L$ is a
non-negative integer. $D_x>D_s$ is the \emph{undercomplete},
$D_x=D_s$ is the \emph{complete}, and $D_x<D_s$ is the
\emph{overcomplete} task. Here, we treat the undercomplete BSSD
(uBSSD) problem.

For consecutive reductional steps we rewrite the BSSD model using
operators. Let
\mbox{$\b{H}[z]:=\sum_{l=0}^{L}\b{H}_{l}z^{-l}\in\R[z]^{D_x\times
D_s}$} denote the $D_x\times D_s$ polynomial matrix corresponding
to the convolutive mixing, in a one-to-one manner. Here, $z$ is
the time-shift operation, that is
\mbox{$(z^{-1}\b{u})(t):=\b{u}(t-1)$}. Now, the BSSD equation
\eqref{eq:obs-1} can be written as
\begin{equation}
    \b{x}=\b{H}[z]\b{s}.
\end{equation}
In the uBSSD task it is assumed that $\b{H}[z]$ has a polynomial matrix left inverse. In other words, there
exists polynomial matrix $\b{W}[z]\in\R[z]^{D_s\times D_x}$ such that $\b{W}[z]\b{H}[z]$ is the identity
mapping. It can be shown \cite{rajagopal03multivariate} that for $D_x>D_s$ such a left inverse exists with probability
1, under mild conditions. The mild condition is as follows: Coefficients of polynomial matrix $\b{H}[z]$, that
is, the random matrix $[\b{H}_0;\ldots;\b{H}_L]$ is drawn from a continuous distribution. For the ISA task it is
supposed that mixing matrix $\b{H}_0\in\R^{D_x\times D_s}$ has full column rank, i.e., its rank is $D_s$.

\subsection{Existing Decomposition Principles in the BSSD Problem Family} \label{sec:BSSD-red-principles}
\begin{figure}%
\centering%
\psset{framesep=0.05,doublesep=0.1}
\xymatrixcolsep{6.5cm} \xymatrixrowsep{0.8cm} 
  \[
  \xymatrix{
  \PstFrameBoxThreeD[FrameBoxThreeDColorHSB=0 0 0.7]{\txt{BSSD}}\myar@{-->}[r]\myar@{-->}[d]\myar@<1.2ex>[r]^(0.45){\text{uBSSD: AR fit (this paper)}}\myar@<-1.3ex>[r]_(0.49){\text{uBSSD: concatenation in time \cite{szabo07undercomplete}}} & \PstFrameBoxThreeD[FrameBoxThreeDColorHSB=0 0 0.7]{\txt{ISA}}\myar@<-0.7ex>@{-->}[d]\myar@<0.7ex>[d]^{\text{Separation Theorem \cite{cardoso98multidimensional,szabo06cross,szabo07undercomplete}}}\\
  \PstFrameBoxThreeD[FrameBoxThreeDColorHSB=0 0 0.7]{\txt{BSD}}\myar@<+0.6ex>[ur]!<-17pt,-16pt>*{} \myar@<+1.2ex>@{}[ur]_(0.11){\begin{rotate}{8}\tiny uBSD: concatenation in time \cite{fevotte03unified}\end{rotate}}
  \myar@{-->}[r]\myar@<-1.3ex>[r]_(0.494){\text{uBSD: spatio-temporal decorrelation \cite{choi97blind}, AR fit \cite{icart96blind,delfosse96adaptive,gorokhov99blind}}} & \PstFrameBoxThreeD[FrameBoxThreeDColorHSB=0 0 0.7]{\txt{ICA}} }
  \]
\caption[]{Extensions of the ICA task. Prefix \emph{u} denotes the
undercomplete case. Dotted arrows point to special cases. Solid
arrows indicate possible reductions. Respective
reduction principles are noted at the arrows.}%
\label{fig:special-BSSD-models}%
\end{figure}
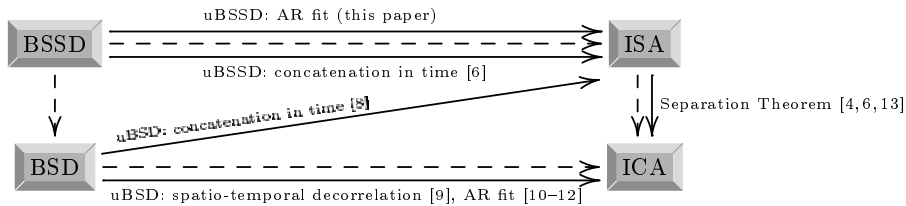

There are numerous reduction methods for the BSSD problem in the
literature. For example, its special case, the undercomplete BSD
task can be reduced (i) to ISA by temporal concatenation of the
observations \cite{fevotte03unified}, or (ii) to ICA by means of
either spatio-temporal decorrelation \cite{choi97blind}, or by linear
prediction (autoregressive (AR) estimation)
\cite{icart96blind,delfosse96adaptive,gorokhov99blind}. As it was
shown in \cite{szabo07undercomplete}, the uBSSD task can also be
reduced to ISA by temporal concatenation. In
Section~\ref{sec:reduction-steps}, we show another route and
describe how linear prediction can help to transcribe the uBSSD
task to ISA. According to the ISA Separation Theorem
\cite{szabo06cross,szabo07undercomplete}, under certain
conditions, the solution of the ISA task requires an ICA
preprocessing step followed by a suitable permutation of the ICA
elements. This principle was conjectured in
\cite{cardoso98multidimensional} on basis of numerical
simulations. Only sufficient conditions are available in
\cite{szabo06cross,szabo07undercomplete} for the ISA Separation
Theorem. Possible reduction steps are shown in
Fig.~\ref{fig:special-BSSD-models}.

\section{Reduction of uBSSD to ISA by Linear Prediction} \label{sec:reduction-steps}
Below, we reduce the uBSSD task to ISA by means of linear prediction. The procedure is similar to that of
\cite{gorokhov99blind}, where it was applied for undercomplete BSD (i.e., for $d=1$).
\begin{mainthm}
In the uBSSD task, observation process $\b{x}(t)$ is
autoregressive and its innovation
$\tilde{\b{x}}(t):=\b{x}(t)-E[\b{x}(t)|\b{x}(t-1),\b{x}(t-2),\ldots]$
is $\b{H}_0\b{s}(t)$, where $E[\cdot| \cdot]$ denotes the
conditional expectation value. Consequently, there is a polynomial
matrix $\b{W}_{\mathrm{AR}}[z]\in\R[z]^{D_x\times D_x}$ such that
$\b{W}_{\mathrm{AR}}[z]\b{x}=\b{H}_0\b{s}$.
\end{mainthm}
\begin{proof}
We assumed that $\b{H}[z]$ has left inverse, thus the hidden $\b{s}$  can be expressed from observation $\b{x}$
by causal FIR filtering, i.e., $\b{s}=\b{H}^{-1}[z]\b{x}$, where
$\b{H}^{-1}[z]=\sum_{n=0}^N\b{M}_nz^{-n}\in\R[z]^{D_s\times D_x}$ and $N$ denotes the degree of the
$\b{H}^{-1}[z]$ polynomial. Thus, terms in observation $\b{x}$ that differ from $\b{H}_0\b{s}(t)$ in
\eqref{eq:obs-1} belong to the linear hull of the finite history of $\b{x}$: $\b{x}(t)=\b{H}_0\b{s}(t) +
\sum_{l=1}^{L}\b{H}_l(\b{H}^{-1}[z]\b{x})(t-l) \in \b{H}_0\b{s}(t) +
\left<\b{x}(t-1),\b{x}(t-2),\ldots,\b{x}(t-L+N)\right>$. Because $\b{s}(t)$ is independent of
$\left<\b{x}(t-1),\b{x}(t-2),\ldots,\b{x}(t-L+N)\right>$, we have that observation process $\b{x}(t)$ is
autoregressive with innovation $\b{H}_0\b{s}(t)$.
\end{proof}
Thus, AR fit of $\b{x}(t)$ can be used for the estimation of
$\b{H}_0\b{s}(t)$. This innovation corresponds to the observation
of an undercomplete ISA model\footnote{Assumptions made for
$\b{H}[z]$ in the uBSSD task implies that $\b{H}_0$ is of full
column rank and thus the resulting ISA task is well defined.},
which can be reduced to a complete ISA using principal component
analysis (PCA). Finally, the solution can be finished by any ISA
procedure. The pseudocode of the above linear predictive
approximation (LPA) method for the uBSSD task is given in
Table~\ref{tab:LPA-pseudocode}.

\begin{table}
  \centering
  \caption{Linear predictive approximation (LPA): Pseudocode} \label{tab:LPA-pseudocode}
  \begin{tabular}{|l|}
        \hline
        \textbf{Input of the algorithm}\\
        \verb|   |Observation: $\{\mathbf{x}(t)\}_{t=1,\ldots,T}$\\
        \textbf{Optimization}\\
        \verb|   |\textbf{AR fit}: for observation $\b{x}$ estimate $\hat{\b{W}}_{\text{AR}}[z]$\\
        \verb|   |\textbf{Estimate innovation}: $\tilde{\b{x}}=\hat{\b{W}}_{\text{AR}}[z]\b{x}$\\
        \verb|   |\textbf{Reduce uISA to ISA and whiten}: $\tilde{\b{x}}^{'}=\hat{\b{W}}_{\text{PCA}}\tilde{\b{x}}$\\
        \verb|   |\textbf{Apply ISA for $\tilde{\b{x}}^{'}$}: separation matrix is $\hat{\b{W}}_{\text{ISA}}$\\
        \textbf{Estimation}\\
        \verb|    |$\hat{\b{W}}_{\text{uBSSD}}[z]=\hat{\b{W}}_{\text{ISA}}\hat{\b{W}}_{\text{PCA}}\hat{\b{W}}_{\text{AR}}[z]$\\
        \verb|               |$\hat{\b{s}}=\hat{\b{W}}_{\text{uBSSD}}[z]\b{x}$\\
        \hline
 \end{tabular}
\end{table}

The reduction procedure implies that hidden components $\b{s}^m$ can be recovered only up to the ambiguities of
the ISA task. The ISA ambiguities are simple \cite{theis04uniqueness1}: hidden multidimensional components can
be determined up to permutation and up to invertible transformation within the subspaces. Furthermore, in the
ISA model it can be assumed without any loss of generality, that both the hidden source ($\b{s}$) and the
observation are white; their expectation values are zeroes and the covariance matrices are identities. Now, the
$\b{s}^m$ components are determined up to permutation and orthogonal transformation.

\section{Illustrations} \label{sec:illustrations}
We show the results of our studies concerning the efficiency of the algorithm of Table~\ref{tab:LPA-pseudocode}. We compare the LPA procedure with the uBSSD method described in
\cite{szabo07undercomplete}. There temporal concatenation was applied to transform the uBSSD task to a
`high-dimensional' ISA task. We shall refer to that method as the method of temporal concatenation, or TCC for
short. Test problems are introduced in Section~\ref{sec:databases}. The performance index that we use to measure
the quality of the solutions is detailed in Section~\ref{sec:amari-index}. Numerical results are presented in
Section~\ref{sec:simulations}.

\subsection{Databases}\label{sec:databases}
\begin{figure}%
\centering%
\subfloat[][(a)]{
\includegraphics[width=1.2cm]{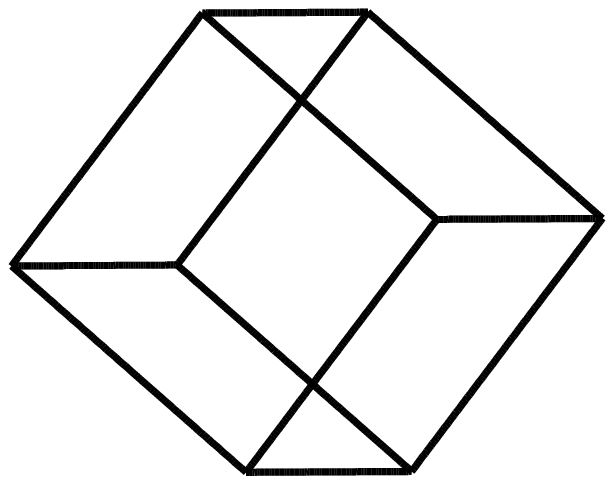}\hfill%
\includegraphics[width=1.2cm]{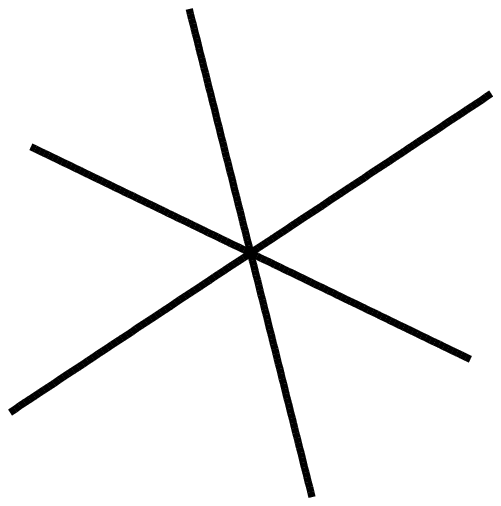}\hfill%
\includegraphics[width=1.2cm]{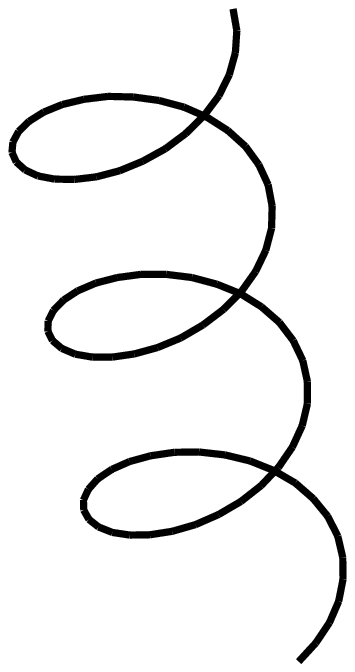}\hfill%
\includegraphics[width=1.2cm]{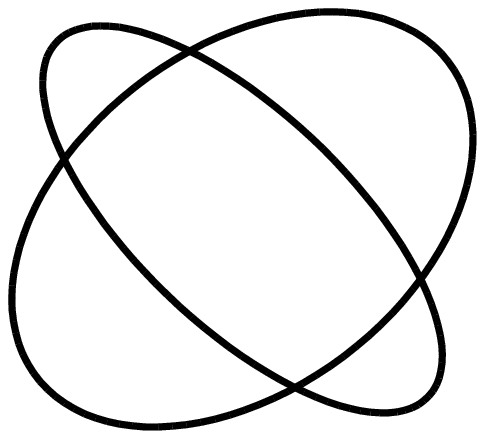}\hfill%
\includegraphics[width=1.2cm]{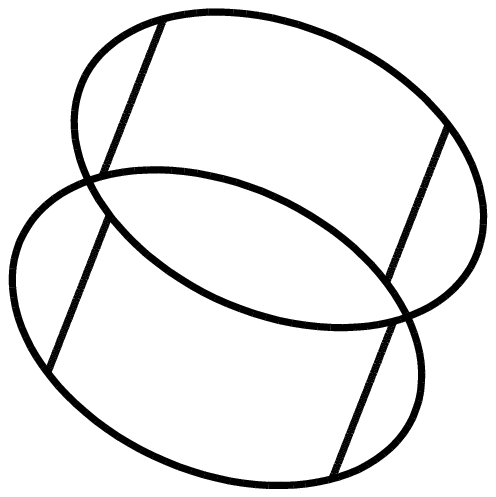}\hfill%
\includegraphics[width=1.2cm]{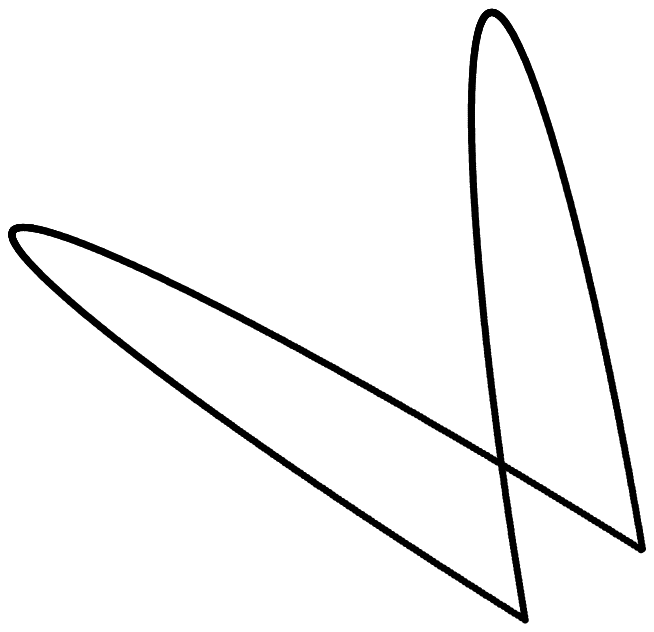}%
}\hfill \subfloat[][(c)]{
\includegraphics[height=1.1cm]{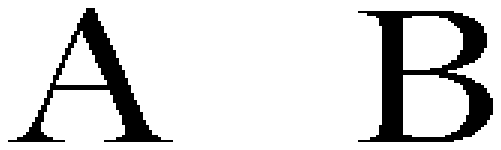}\hfill
}\\
\subfloat[][(b)]{
\includegraphics[width=1.13cm]{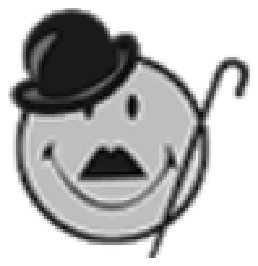}
\includegraphics[width=1.13cm]{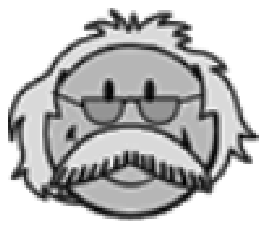}
\includegraphics[width=1.13cm]{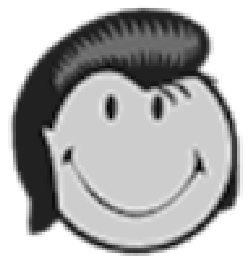}
\includegraphics[width=1.13cm]{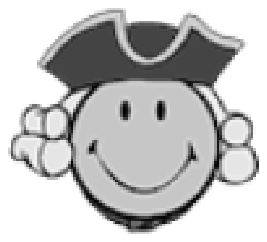}
\includegraphics[width=1.13cm]{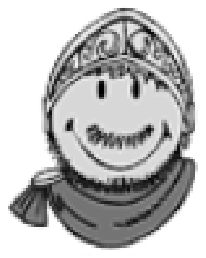}
\includegraphics[width=1.13cm]{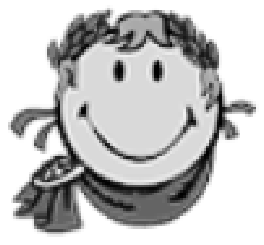}
\includegraphics[width=1.13cm]{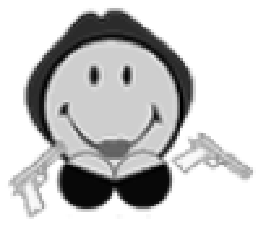}
\includegraphics[width=1.13cm]{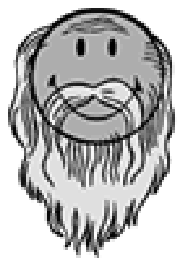}
\includegraphics[width=1.13cm]{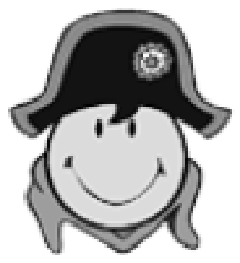}
\includegraphics[width=1.13cm]{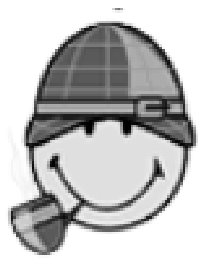}
} \label{fig:database-ABC} \caption[]{Illustration of the
\emph{3D-geom}, \emph{celebrities} and \emph{letters} databases.
(a): Database \emph{3D-geom} contains 6 of 3-dimensional
components ($M=6$, $d=3$). Hidden sources are uniformly
distributed variables on 3-dimensional geometric objects. (b):
Database \emph{celebrities} contains 10 of 2-dimensional
components ($M=10$, $d=2$). Density functions of the hidden
sources ($\b{s}^m$) are proportional to the pixel intensities of
the 2-dimensional images. (c): \emph{Letters} database is
minimal. Hidden sources $\b{s}^m$ are uniformly distributed on images of letters A and B ($M=2$, $d=2$).}%
\label{fig:databases}%
\end{figure}

We define four databases ($\b{s}$) to study our LPA algorithm. These are the databases used in
\cite{szabo07undercomplete}, too. In the \emph{3D-geom} test hidden components $\b{s}^m$ are random variables
uniformly distributed on \mbox{3-dimensional} geometric forms ($d=3$). We have 6 components ($M=6$). The
dimension of the hidden source $\b{s}$ is $D_s=18$. See Fig.~\ref{fig:databases}(a). The \emph{celebrities} test
has 10 of 2-dimensional source components generated from cartoons of celebrities
\mbox{($d=2$)}.\footnote{http://www.smileyworld.com} The 2-dimensional images of celebrities are considered as
the density functions of $\b{s}^m$: sources are generated according to the pixel intensities. See
Fig.~\ref{fig:databases}(b). In the \emph{letters} data set, hidden sources $\b{s}^m$ are uniformly distributed
on \mbox{2-dimensional} images ($d=2$) of letters A and B. The number of components and the dimension of the
sources are minimal ($M=2$, $D_s=4$). See Fig.~\ref{fig:databases}(c).  Our \emph{Beatles} test is a non-i.i.d.\
example. Here, hidden sources are stereo Beatles songs.\footnote{http://rock.mididb.com/beatles/} $8$ kHz
sampled portions of two songs (A Hard Day's Night, Can't Buy Me Love) made the hidden $\b{s}^m$s
($d=2,M=2,D_s=4$).

\captionsetup[subfloat]{labelformat=empty} 
\captionsetup[subfloat]{labelformat=parens}

\subsection{The Amari-index}\label{sec:amari-index}
According to Section~\ref{sec:reduction-steps}, in the ideal case, the product of matrix
$\hat{\b{W}}_{\text{ISA}}\hat{\b{W}}_{\text{PCA}}$ (the result of PCA and ISA) and matrix $\b{H}_0$, that is matrix
$\b{G}:=\hat{\b{W}}_{\text{ISA}}\hat{\b{W}}_{\text{PCA}}\b{H}_0\in\R^{D_s\times D_s}$ is a block-permutation matrix
made of $d\times d$ blocks. To measure this block-permutation property, we used the normalized version
\cite{szabo06cross} of the Amari-error \cite{amari96new} adapted to the ISA task \cite{theis05blind}. Namely, let
matrix $\b{G}\in\R^{D_s\times D_s}$ be decomposed into $d\times d$ blocks:
$\b{G}=\left[\b{G}^{i,j}\right]_{i,j=1,\ldots,M}$. Let $g^{i,j}$ denote the sum of the absolute values of the elements
of matrix $\b{G}^{i,j}\in\R^{d\times d}$. Now, the normalized Amari-error, the Amari-index
($r(\cdot)=r_{d,D_s}(\cdot)$) is defined as:
\[
    r(\b{G}):=\frac{1}{2M(M-1)}\left[\sum_{i=1}^M\left(\frac{
\sum_{j=1}^Mg^{i,j}}{\max_jg^{i,j}}-1\right)+
\sum_{j=1}^M\left(\frac{
\sum_{i=1}^Mg^{i,j}}{\max_ig^{i,j}}-1\right)\right].
\]
For matrix $\b{G}$ we have that $0\le r(\b{G})\le 1$. $r(\b{G})=0$
if, and only if $\b{G}$ is a block-permutation matrix with
$d\times d$ sized blocks. Thus, $r(\b{G})=0$ for a perfect
$\b{G}$, whereas in the worst case $r(\b{G})=1$. Given that index
$r$ takes values in $[0,1]$ \emph{independently from} $d$
\emph{and} $D_s$, we can use this measure to compare the TCC and
LPA techniques.

\subsection{Simulations}\label{sec:simulations}
Results on databases \emph{3D-geom}, \emph{celebrities},
\emph{letters} and \emph{Beatles} are provided here. The
experimental studies concern two questions:
\begin{enumerate}
    \item
        The TCC and the LPA methods are compared on uBSSD tasks.
    \item
        The performance as a function of convolution length is studied for the LPA technique.
\end{enumerate}

Our test databases correspond to those of \cite{szabo07undercomplete} and here, we study the \mbox{$D_x=2D_s$}
case, like in the cited reference. Both the TCC and the LPA method reduce the uBSSD task to ISA problems and we
use the Amari-index (Section~\ref{sec:amari-index}) to measure and compare their performances. For all values of
the parameters (sample number: $T$, convolution length: $L+1$), we have averaged the performances upon 50 random
initializations of $\b{s}$ and $\b{H}[z]$. The coordinates of matrices $\b{H}_l$ were chosen independently from
standard normal distribution. We used the Schwarz's Bayesian Criterion \cite{neumaier01estimation} to determine
the optimal order of the AR process. The criterion was constrained: the order $Q$ of the estimated AR process
(see Table~\ref{tab:LPA-pseudocode}) was limited from above, the upper limit was set to twice the length of the
convolution, i.e., $Q\le 2(L+1)$. The AR process was then estimated by the method detailed in
\cite{neumaier01estimation} and \cite{tapio01algorithm}. Both in the case of TCC and in the case of LPA, ISA was
accomplished by joint f-decorrelation (JFD) as detailed in \cite{szabo06real}.

\begin{figure}%
\centering%
\subfloat[][]{\includegraphics[width=6cm]{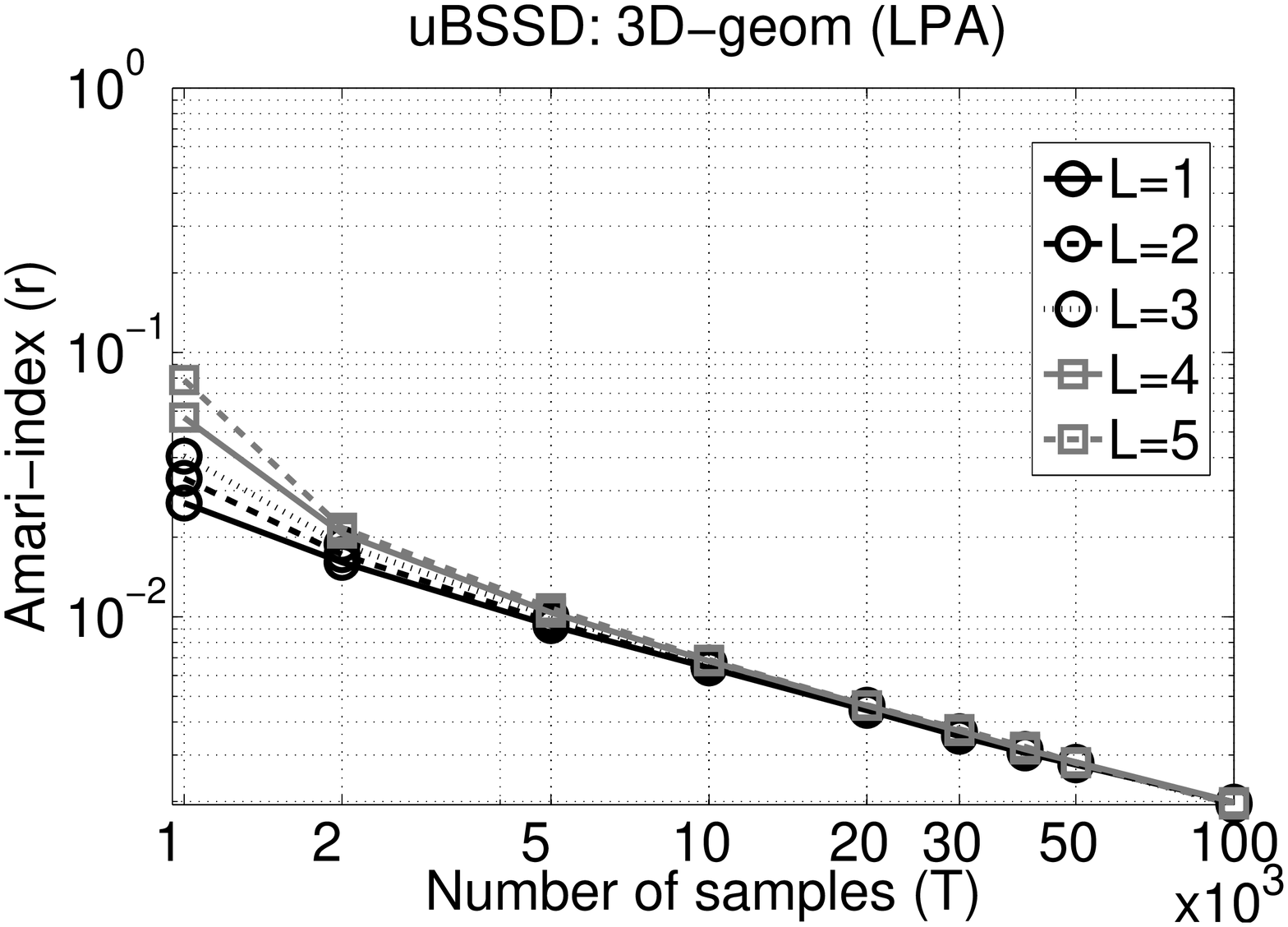}}%
\subfloat[][]{\includegraphics[width=6cm]{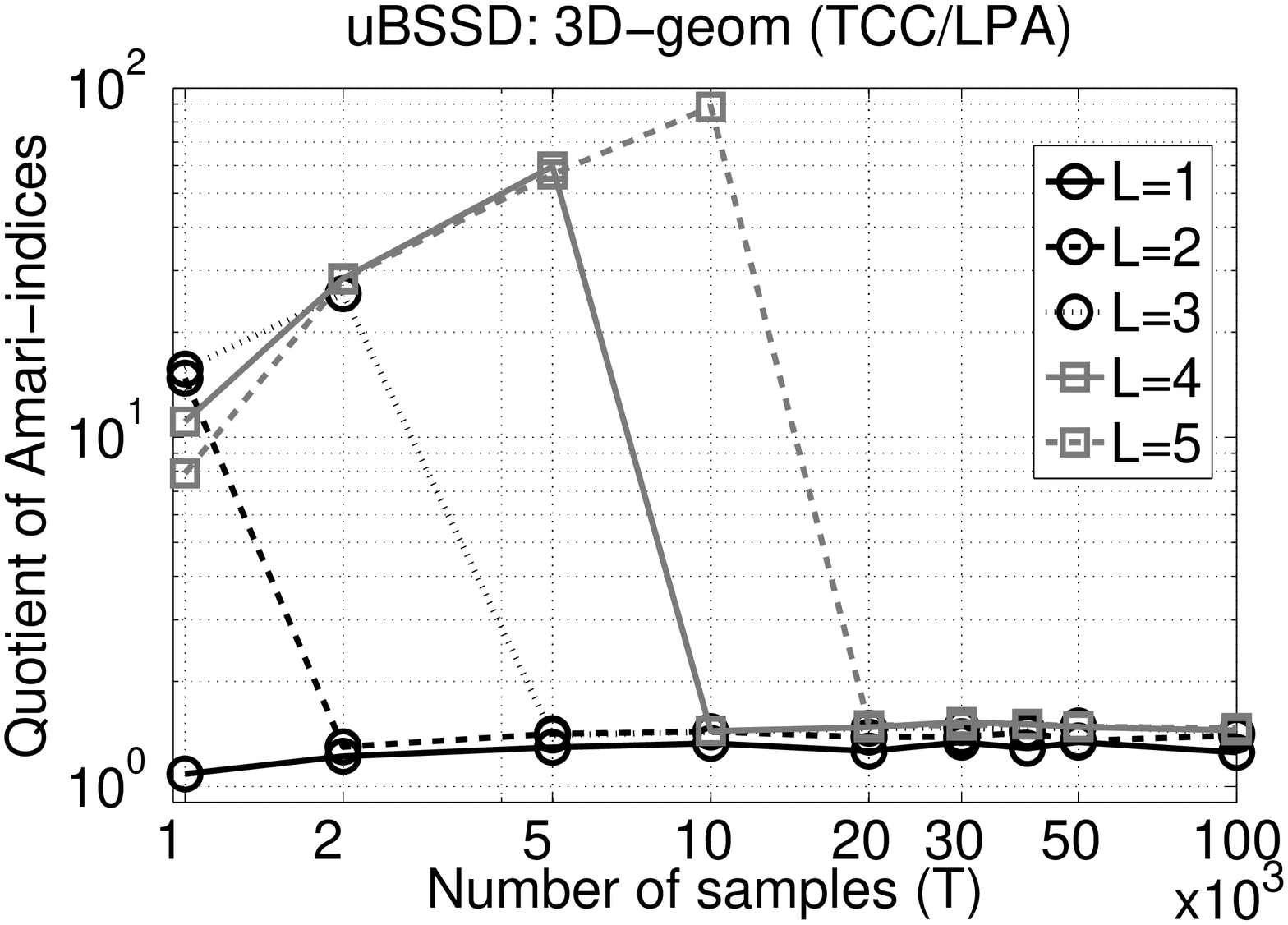}}\\
\subfloat[][]{\includegraphics[width=6cm]{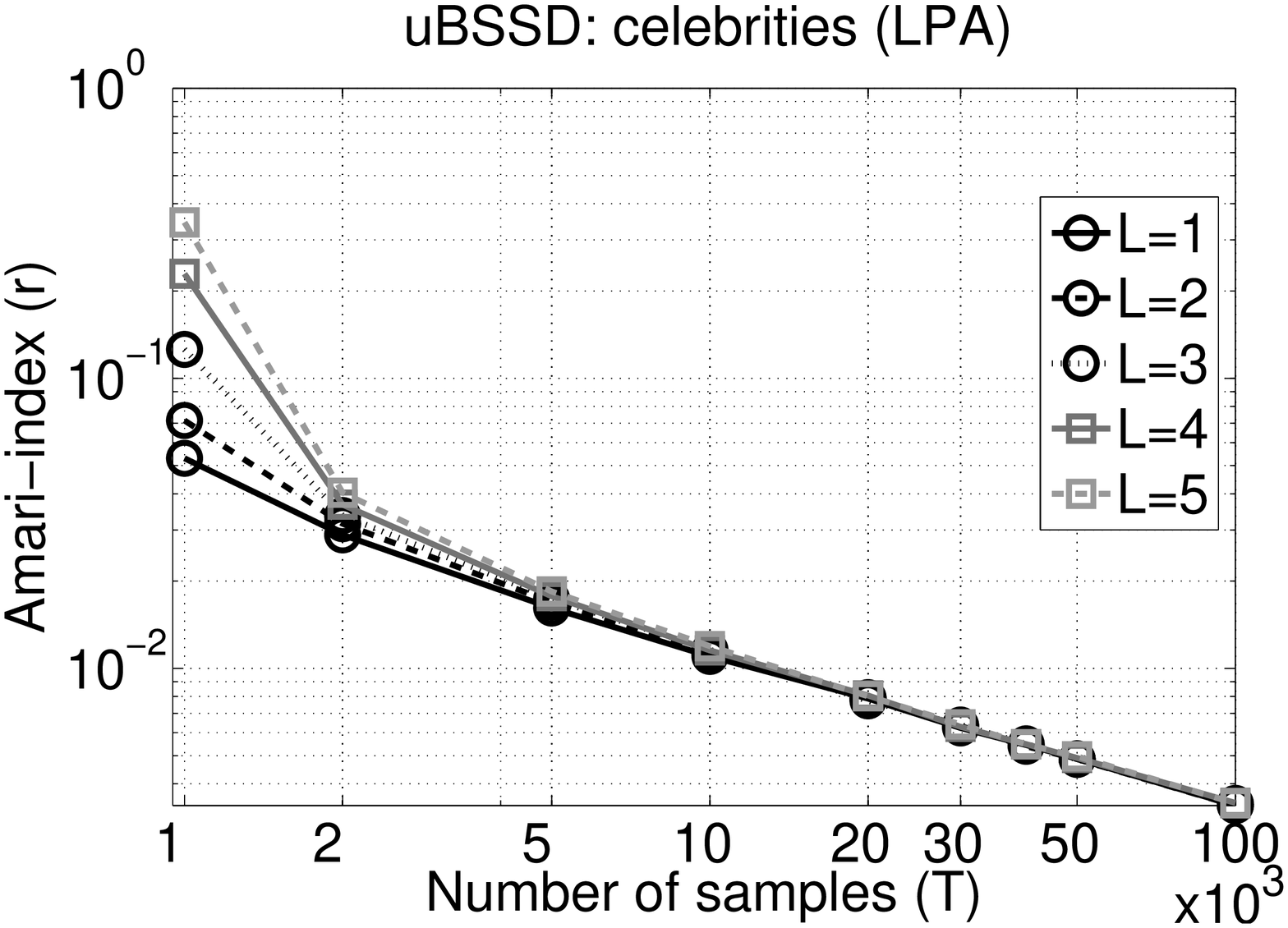}}%
\subfloat[][]{\includegraphics[width=6cm]{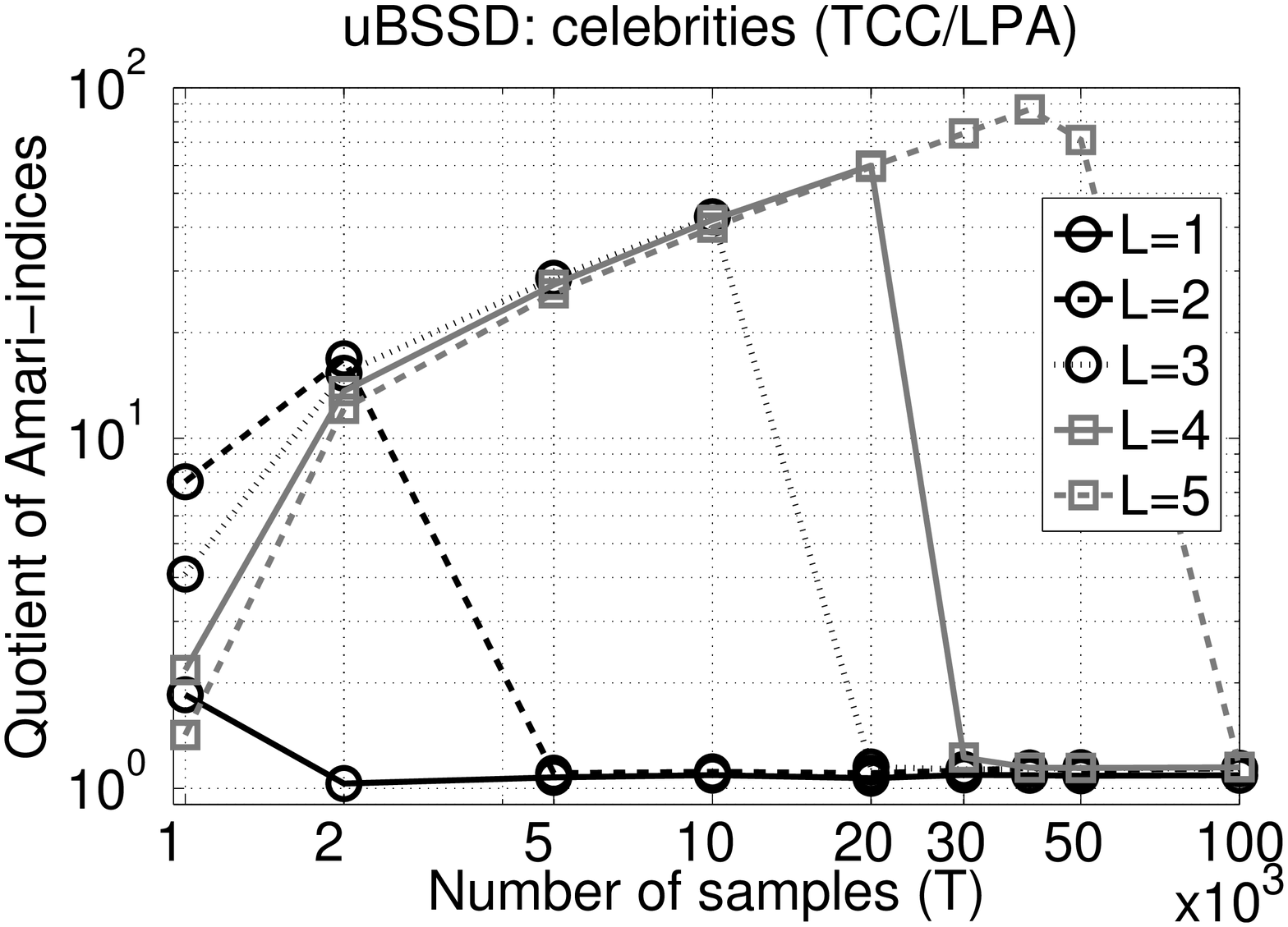}}\\
\subfloat[][]{\includegraphics[width=6cm]{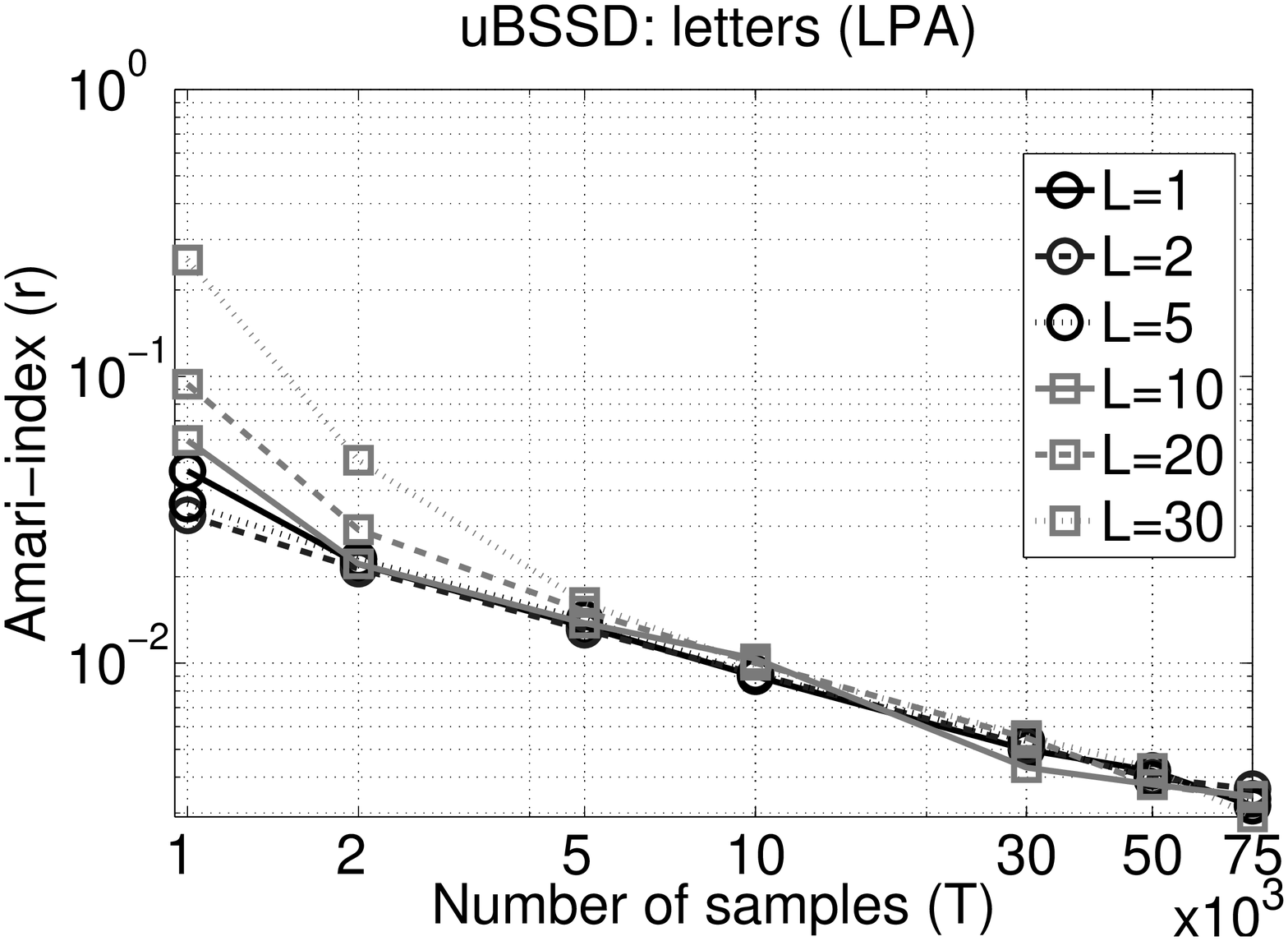}}%
\subfloat[][]{\includegraphics[width=6cm]{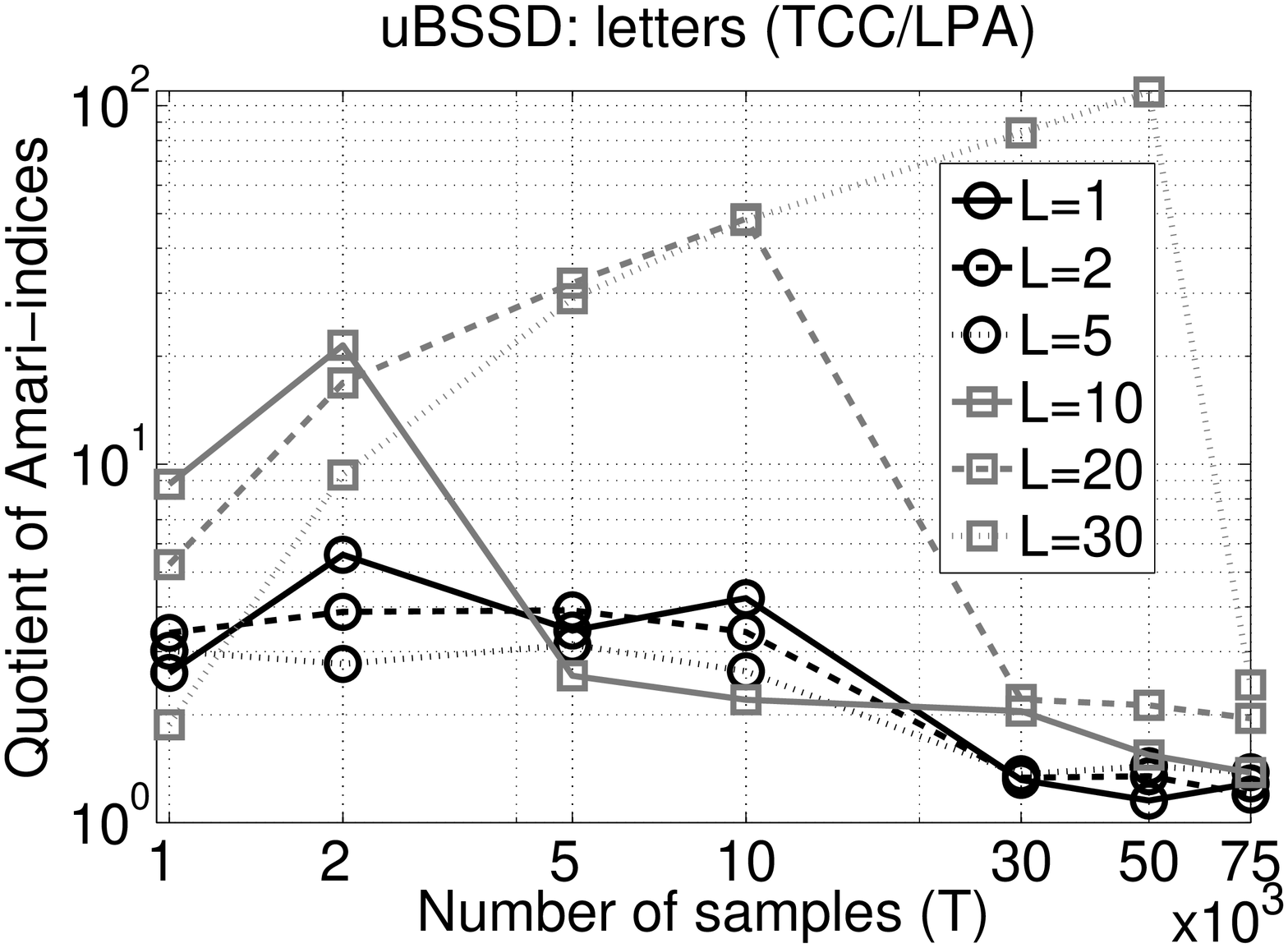}}\\
\subfloat[][]{\includegraphics[width=6cm]{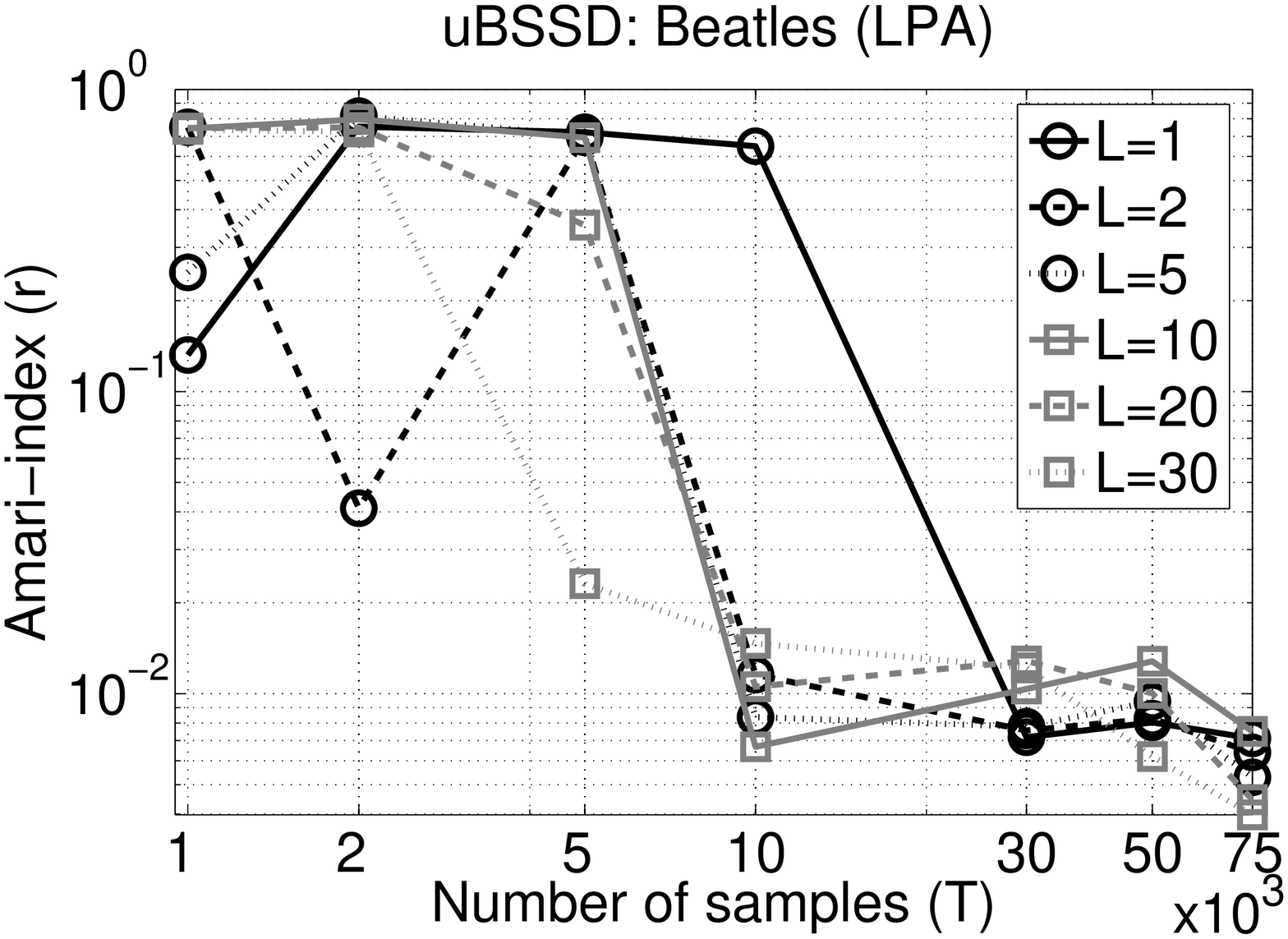}}%
\subfloat[][]{\includegraphics[width=6cm]{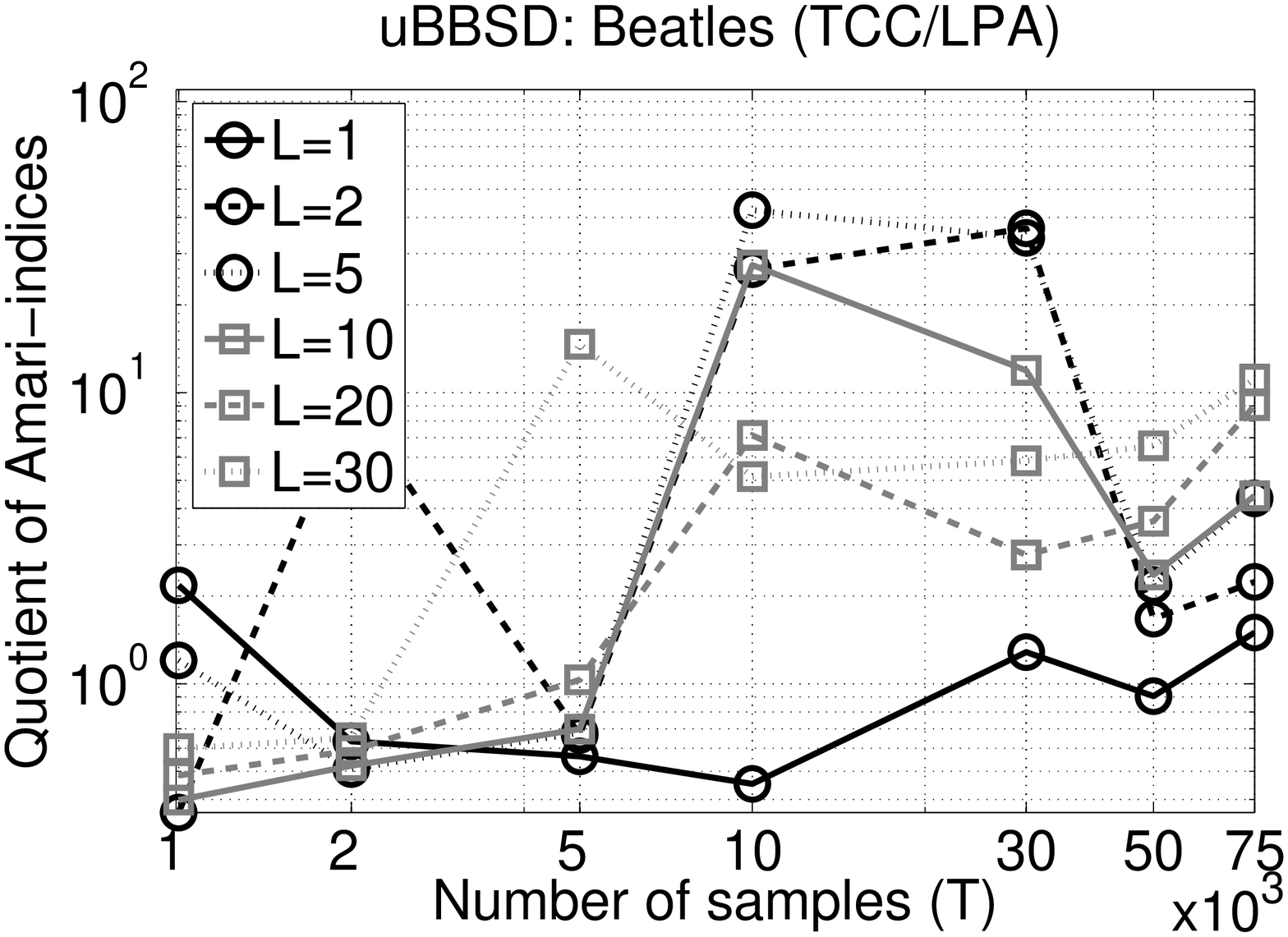}}
\caption[]{Estimation error of the LPA method and comparisons with
the TCC method for the \emph{3D-geom}, the \emph{celebrities}, the
\emph{letters} and the \emph{Beatles} databases. Scales are
`$\log\log$' plots. Data correspond to different convolution
lengths ($L+1$). (a), (c), (e) and (g): Amari-index as a function
of sample number for the \emph{3D-geom}, \emph{celebrities},
\emph{letters} and \emph{Beatles} databases. (b), (d), (f) and
(h): Quotients of the Amari-indices of the TCC and
the LPA methods: for quotient value $q>1$, the LPA method is $q$ times more precise than the TCC method.}%
\label{fig:LPA-r-3D-geom-celebs}%
\end{figure}

\begin{figure}%
\centering%
\subfloat[][]{\includegraphics[height=3.5cm]{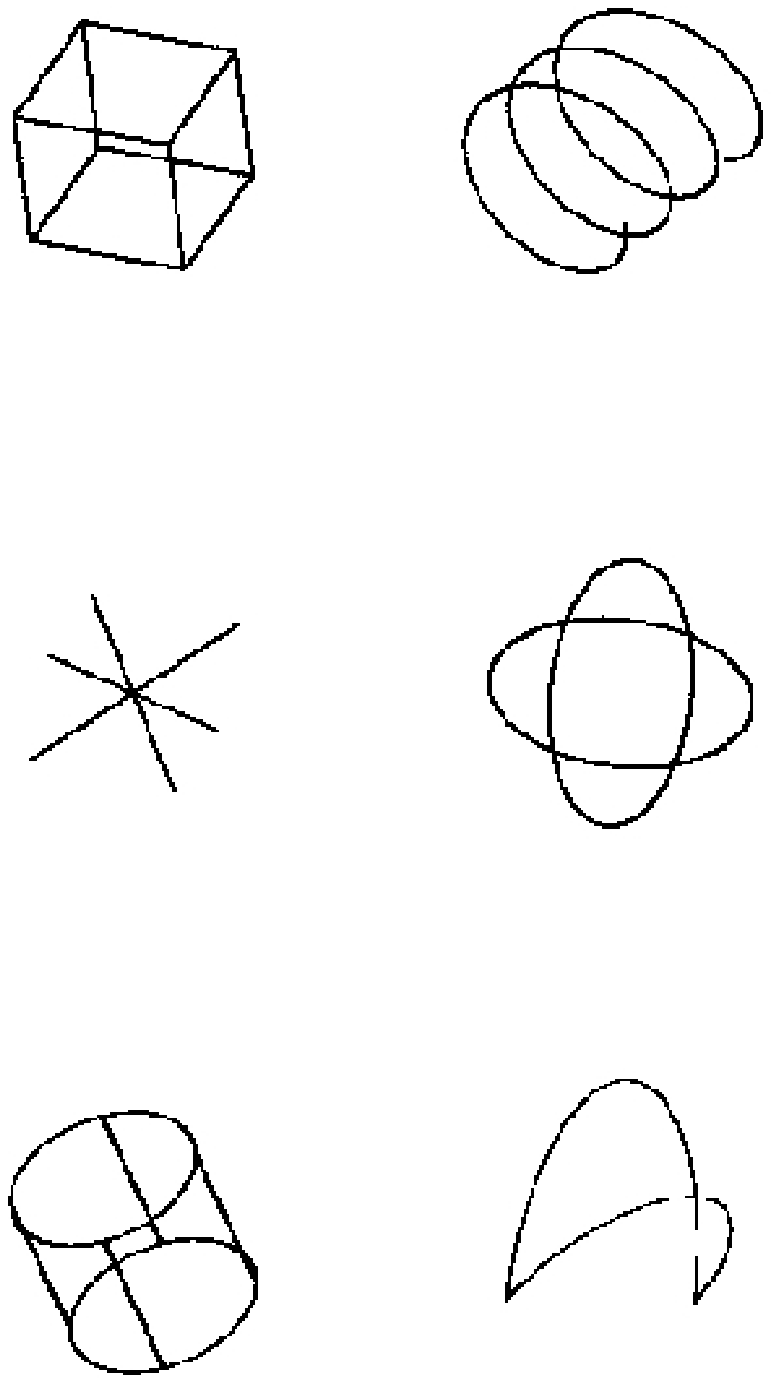}}\hfill%
\subfloat[][]{\includegraphics[height=3.5cm]{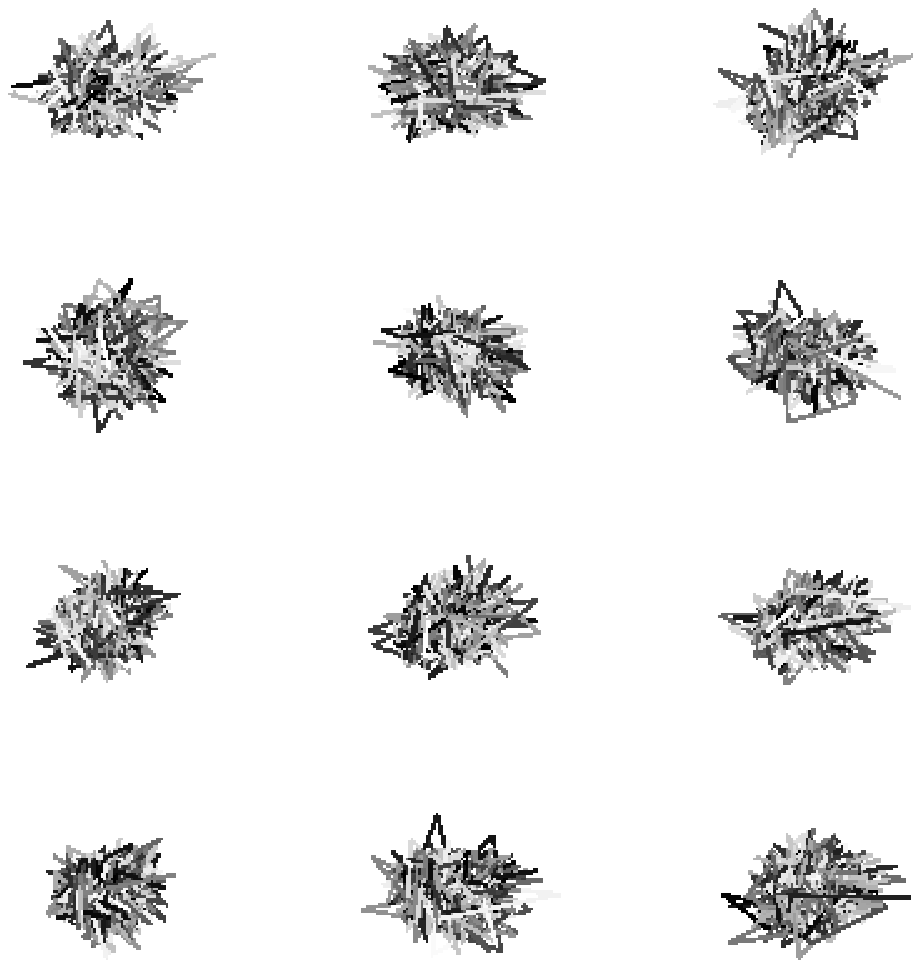}}\hfill%
\subfloat[][]{\includegraphics[height=3.2cm]{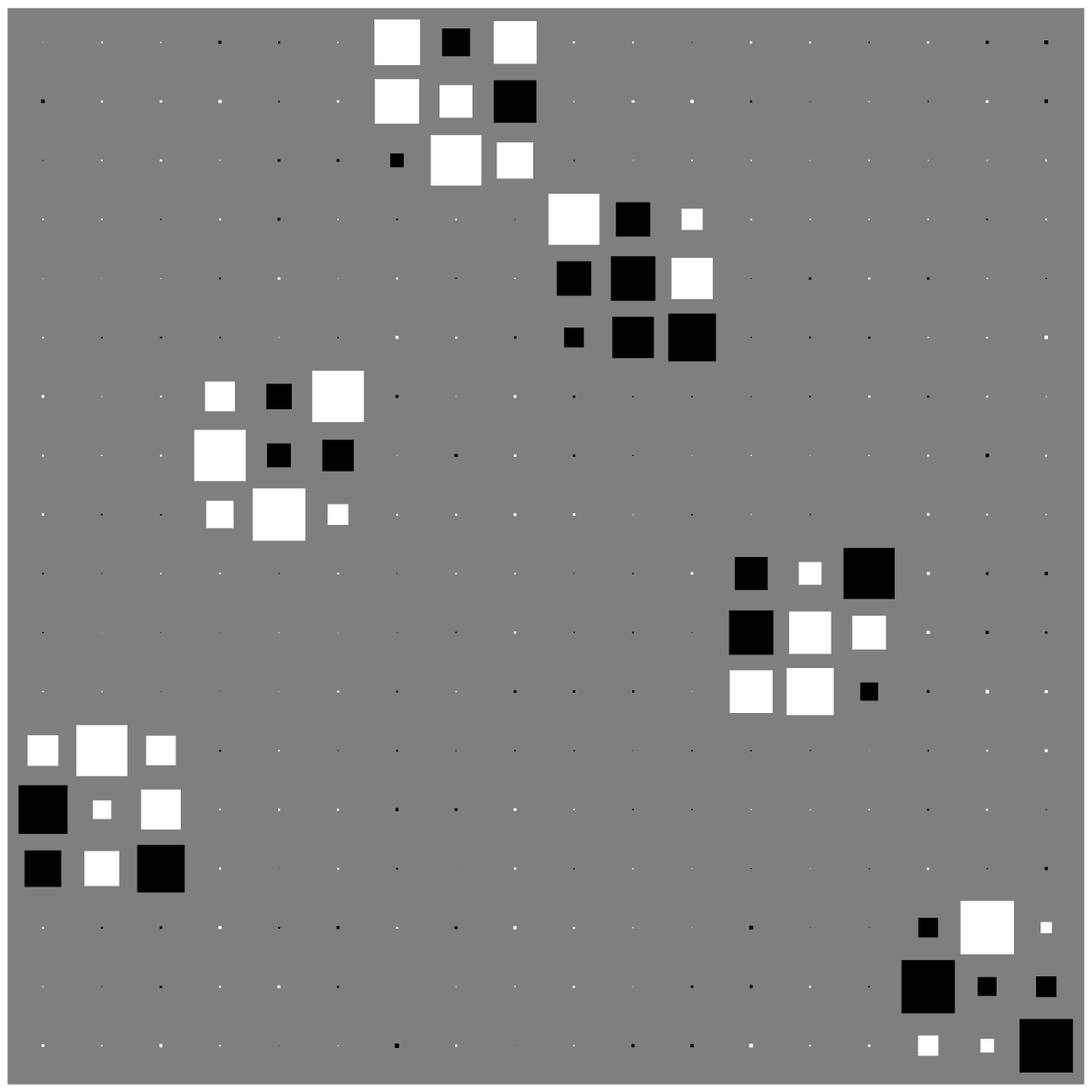}}\hfill%
\subfloat[][]{\includegraphics[height=3.5cm]{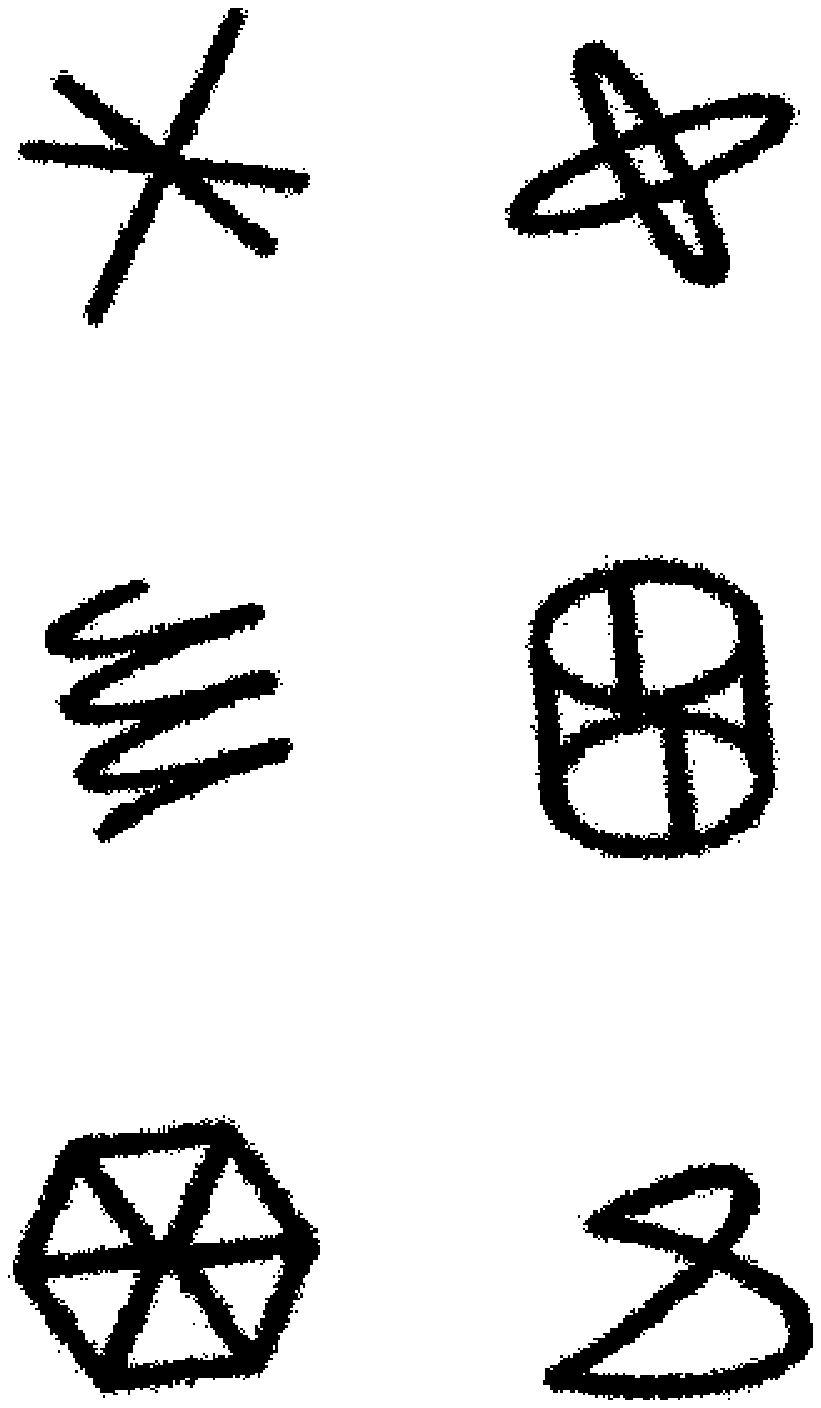}}\\%
\subfloat[][]{\includegraphics[height=3.5cm]{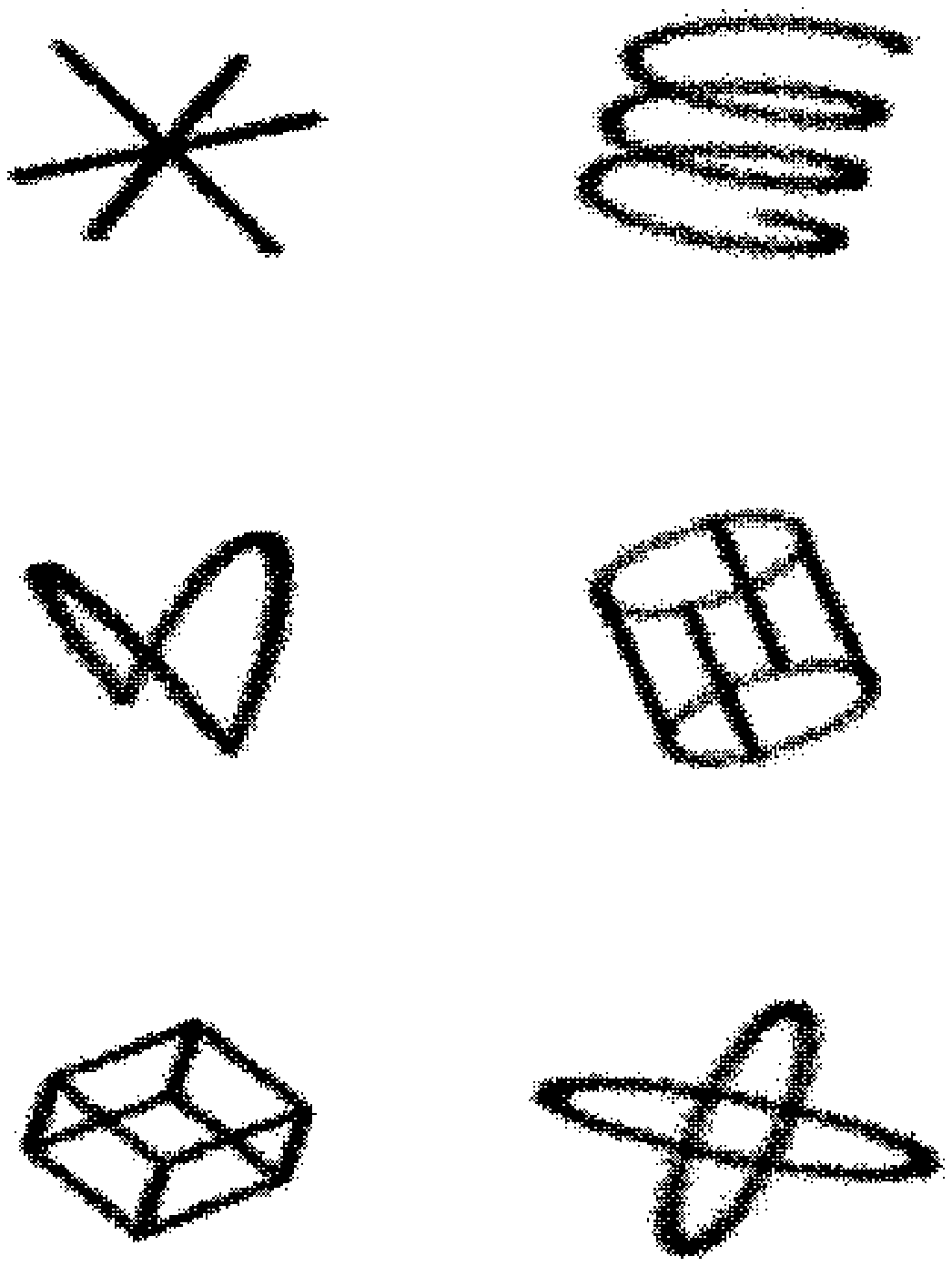}}\hfill%
\subfloat[][]{\includegraphics[height=3.5cm]{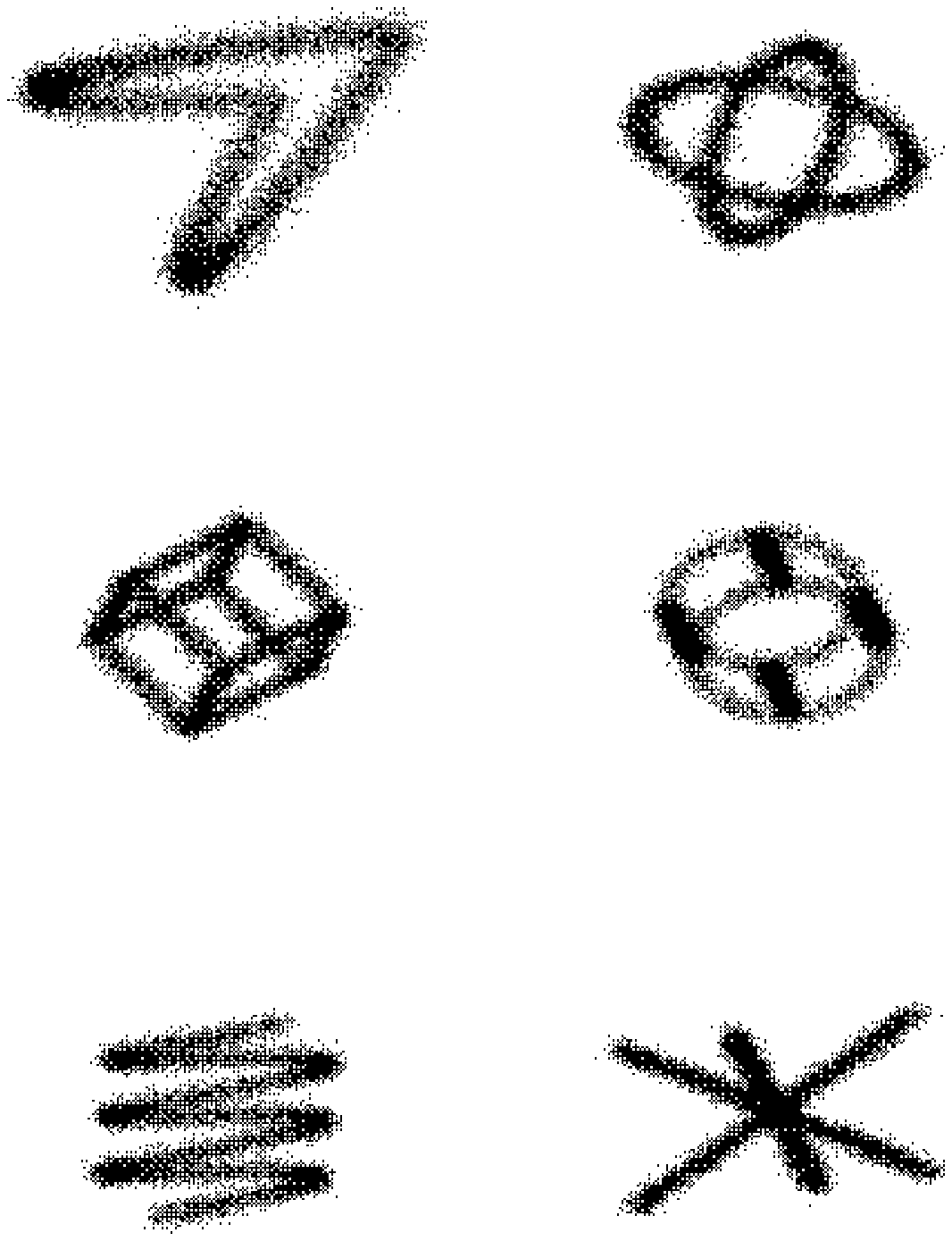}}\hfill%
\subfloat[][]{\includegraphics[height=3.5cm]{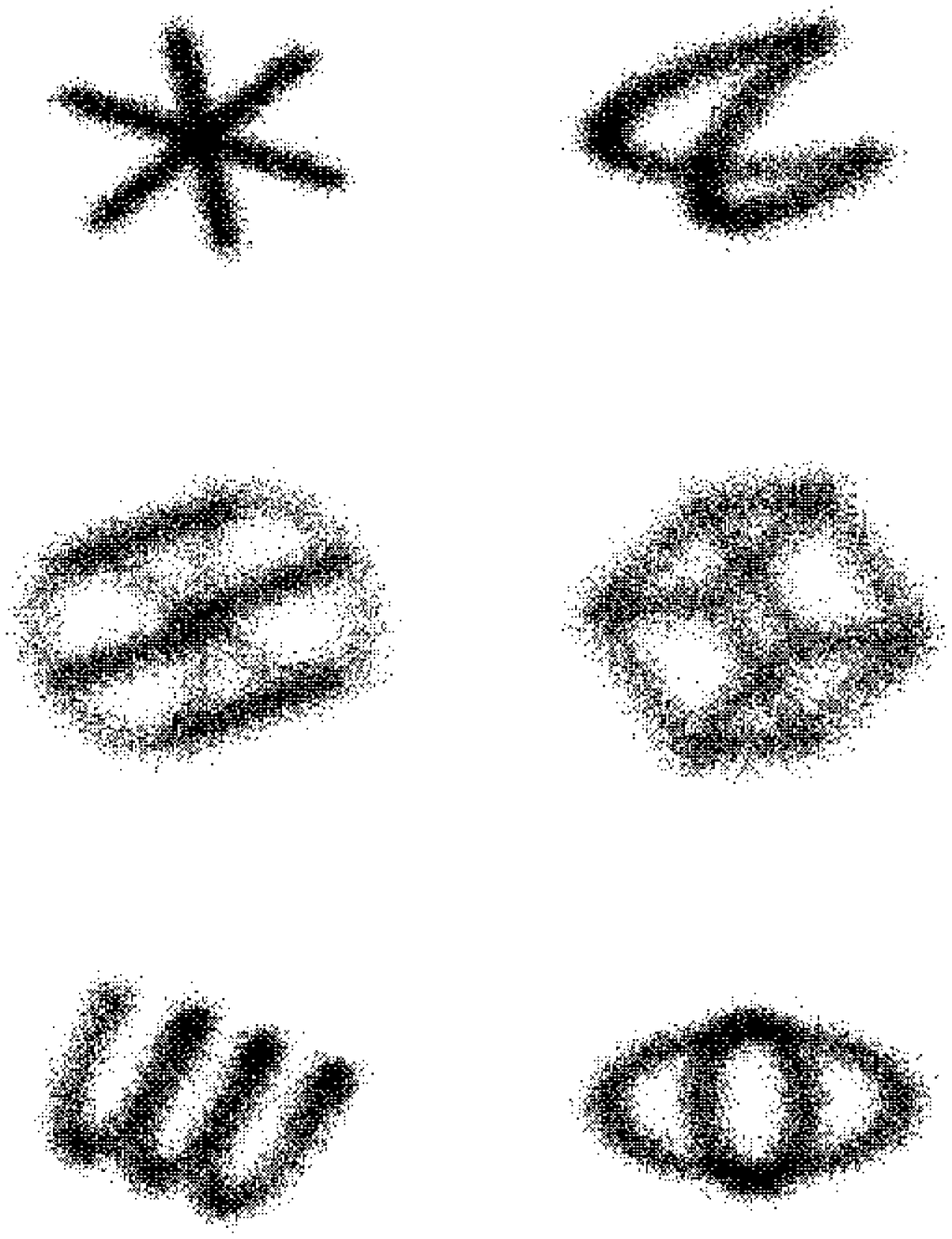}}\hfill%
\subfloat[][]{\includegraphics[height=3.5cm]{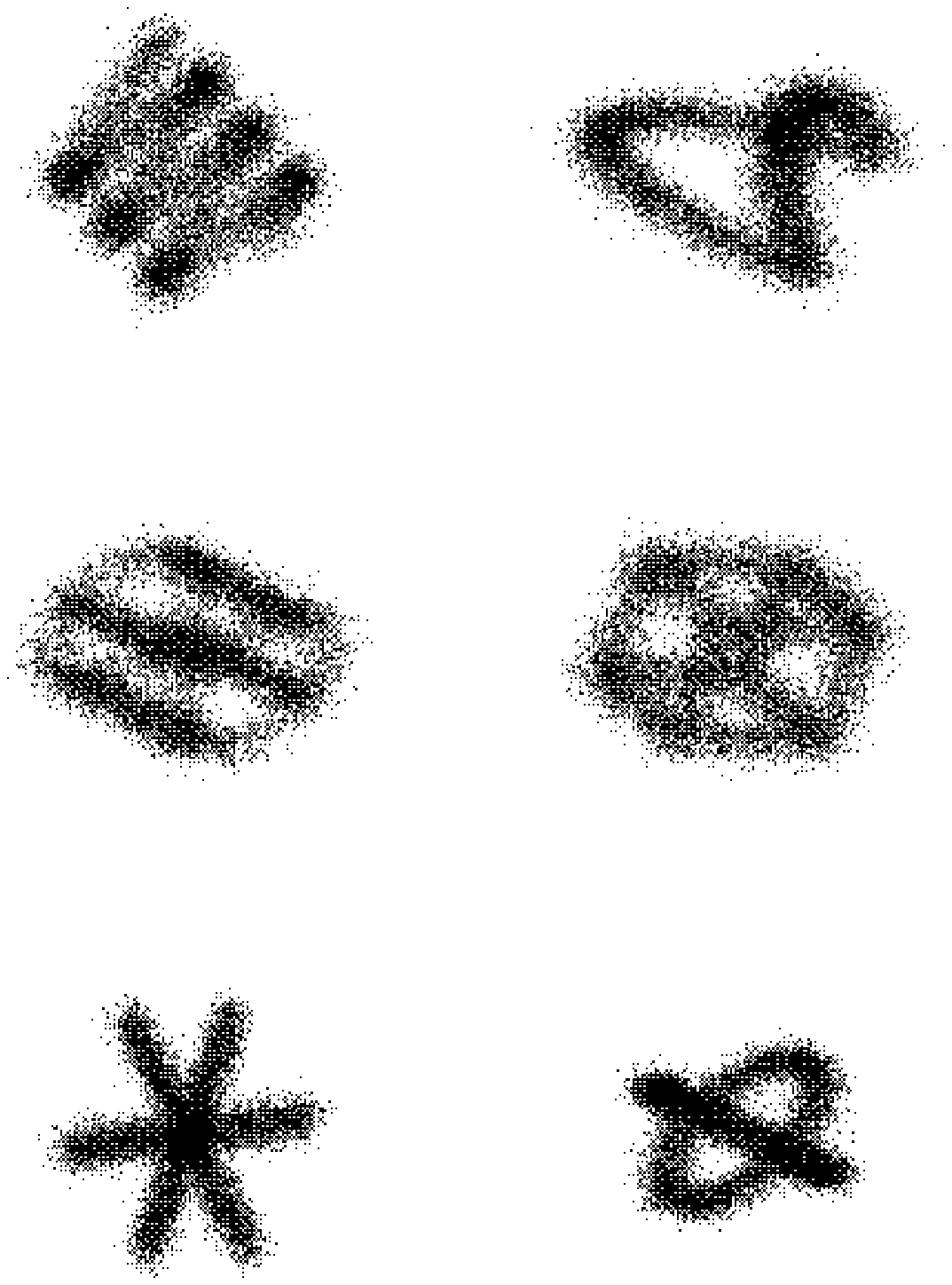}}\\%
\subfloat[][]{\includegraphics[height=3.5cm]{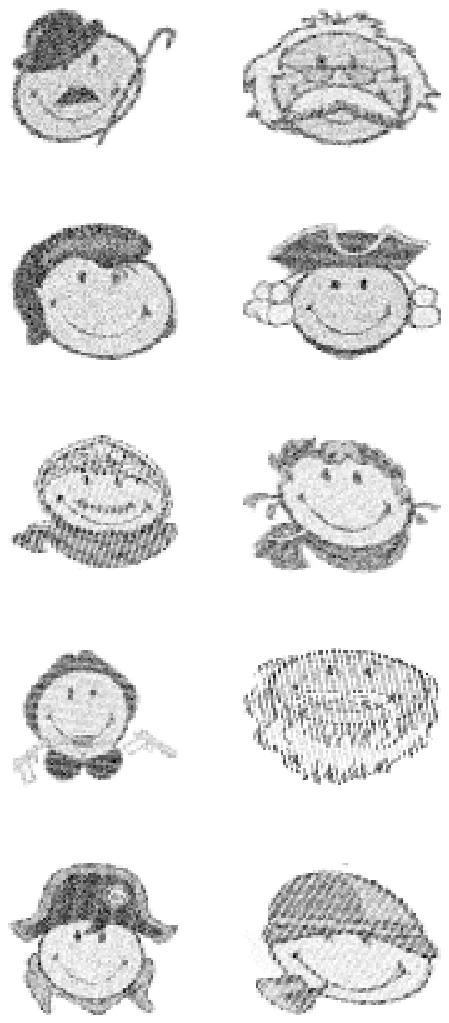}}\hfill%
\subfloat[][]{\includegraphics[height=3.5cm]{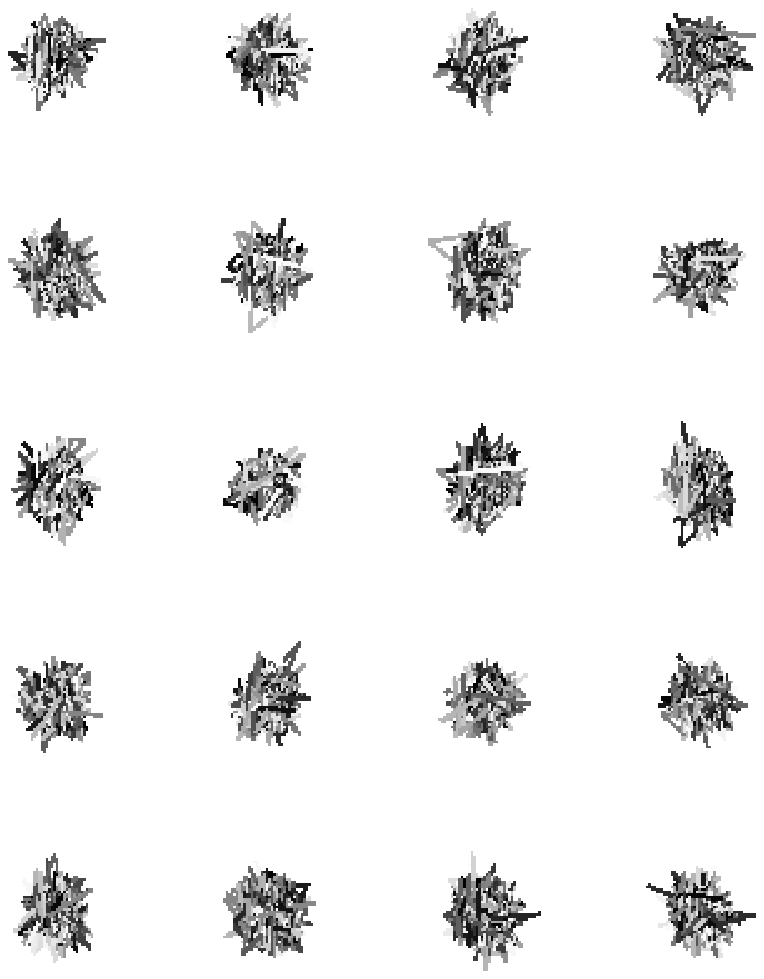}}\hfill%
\subfloat[][]{\includegraphics[height=3.3cm]{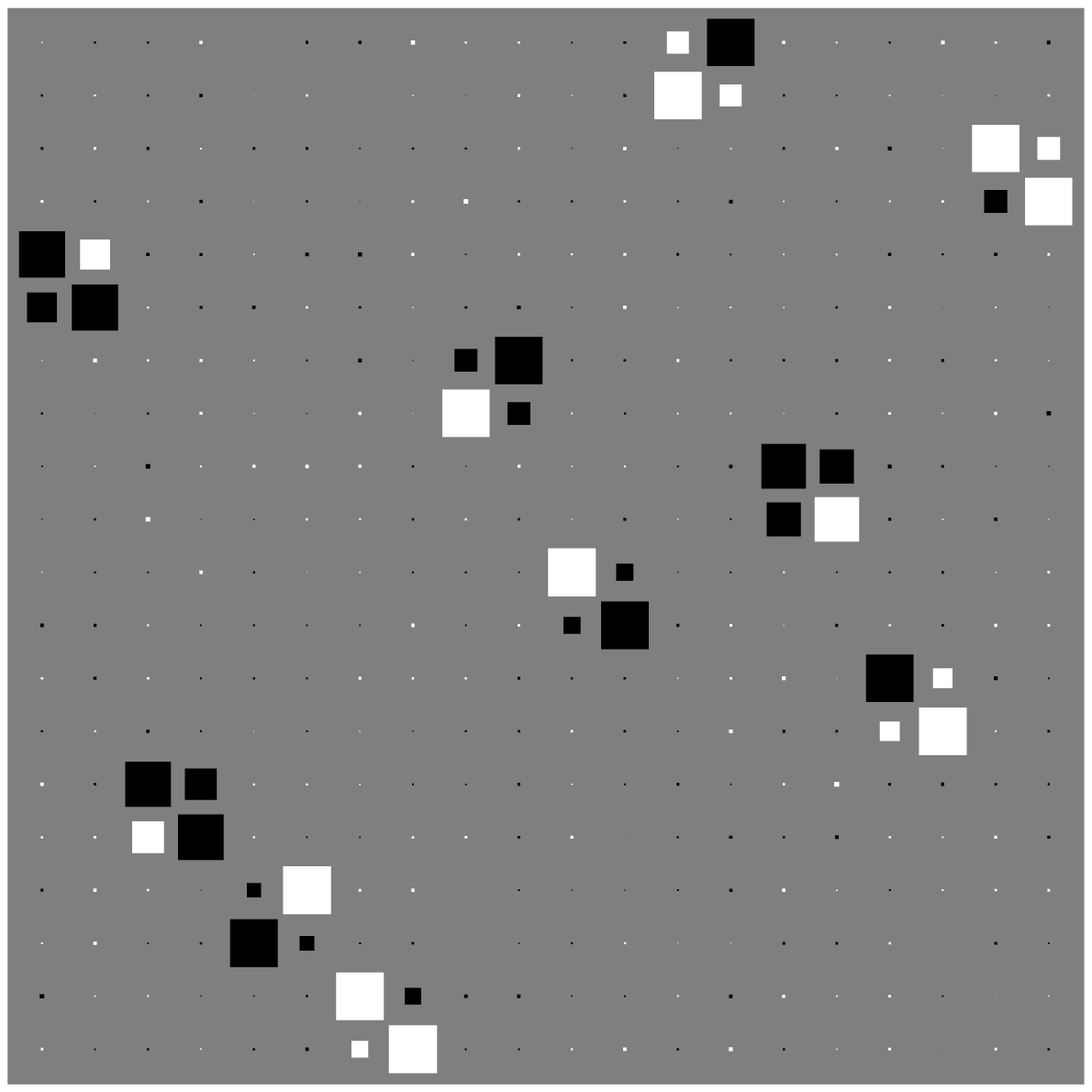}}\hfill%
\subfloat[][]{\includegraphics[height=3.5cm]{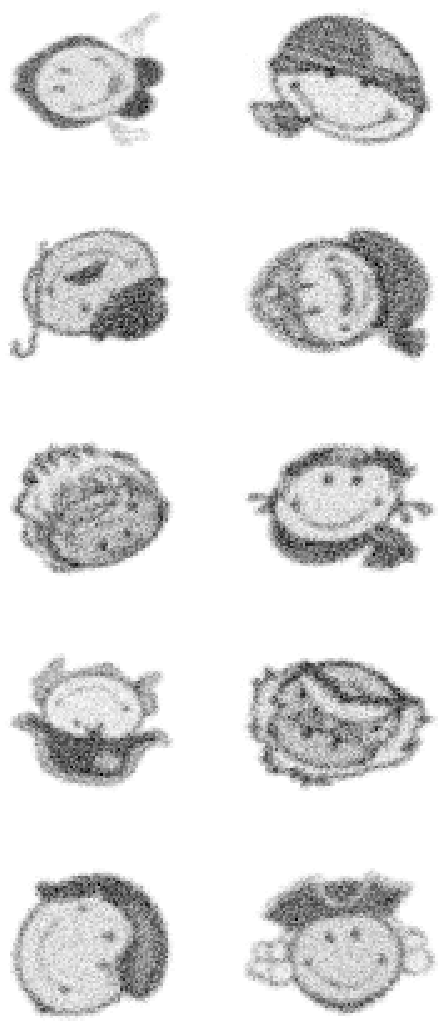}}\\%
\subfloat[][]{\includegraphics[width=2cm]{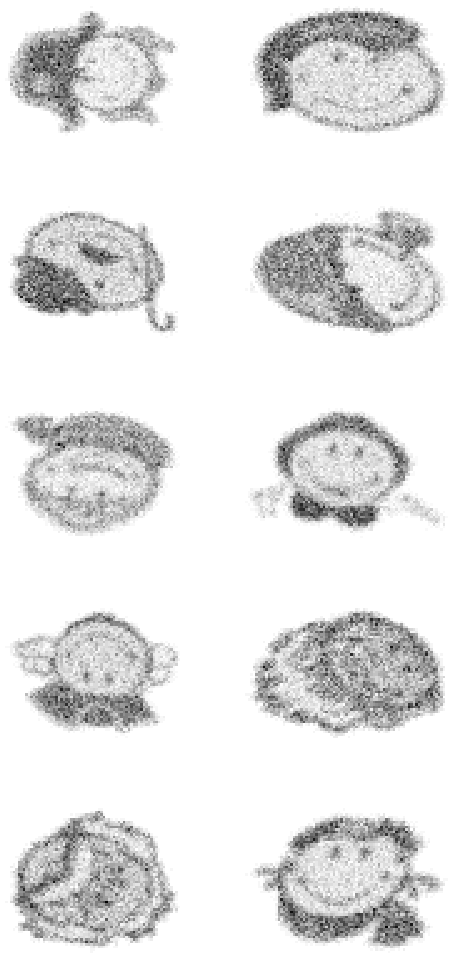}}\hfill%
\subfloat[][]{\includegraphics[width=2cm]{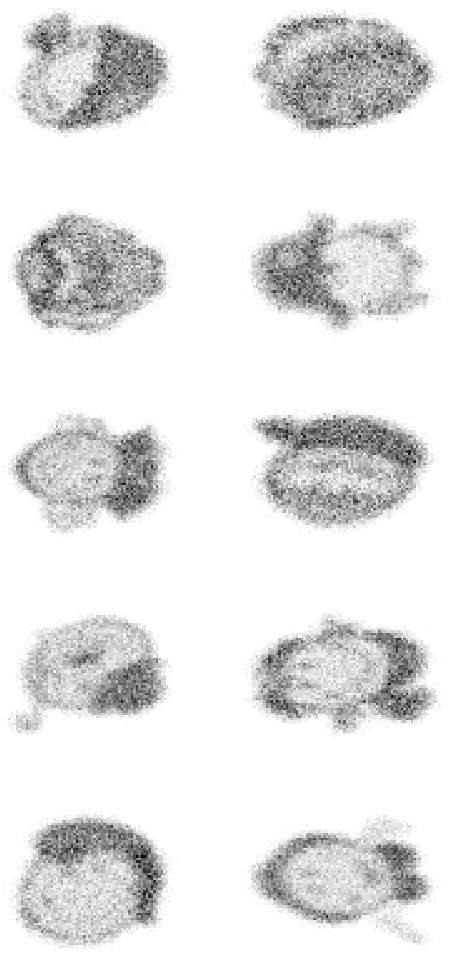}}\hfill%
\subfloat[][]{\includegraphics[width=2cm]{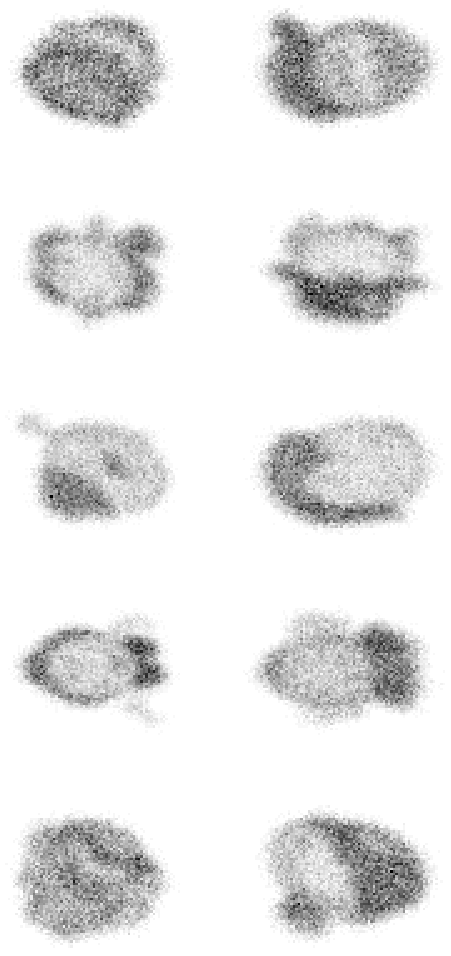}}\hfill%
\subfloat[][]{\includegraphics[width=2cm]{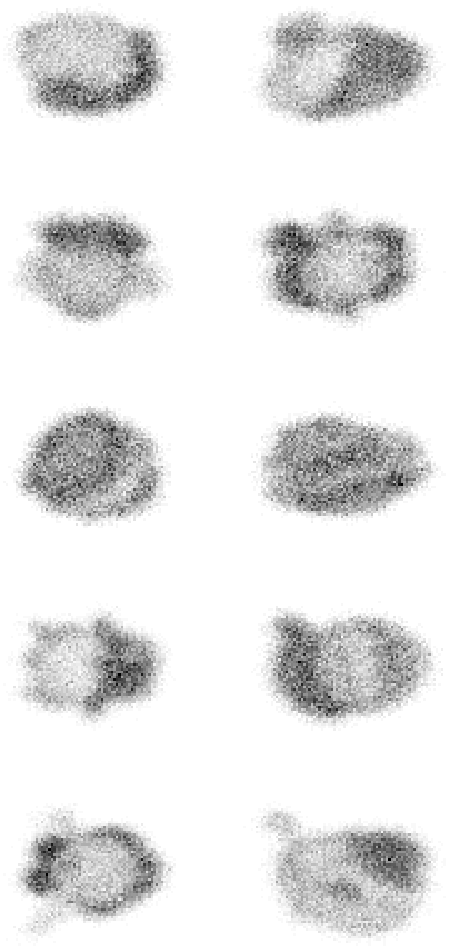}}\hfill%
\subfloat[][]{\includegraphics[width=2cm]{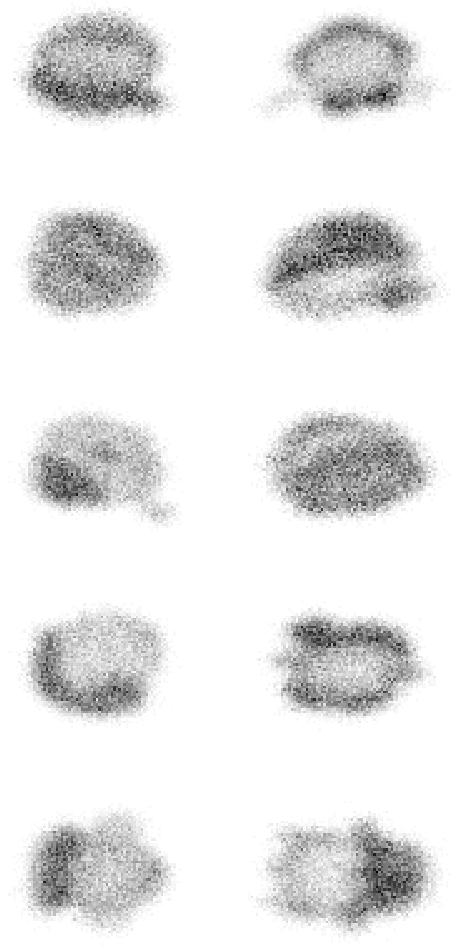}} \caption[]{Illustration of
the LPA method on the uBSSD task for the \emph{3D-geom} and \emph{celebrities} databases. (a)-(d), (i)-(l):
sample number $T=100,000$, convolution length $L+1=6$. (a) and (i): hidden components $\b{s}^m(t)$. (b) and (j):
observed convolved signals $\b{x}(t)$, only $1000$ time steps are shown. (c) and (k): Hinton-diagram of $\b{G}$,
ideally block-permutation matrix with $2\times 2$ ($3\times 3$) blocks. (d) and (l): estimated components
($\hat{\b{s}}^m$), Amari-indices: $0.2\%$ and $0.34\%$, respectively. (e)-(h) and (m)-(q): sample number
$T=20,000$, dependence of estimated components ($\hat{\b{s}}^m$) on the convolution parameter
$L$. $L$ is $1,5,10,20$, and $1,5,10,15,20$ respectively.}%
\label{fig:LPA-demo-3D-geom}%
\end{figure}

We studied the dependence of the precision versus the sample number on databases \mbox{\emph{3D-geom}} and
\emph{celebrities}. The dimension and the number of the components were $d=3$ and $M=6$ for the \mbox{\emph{3D-geom}}
database and $d=2$ and $M=10$ for the \emph{celebrities} database, respectively. In both cases the sample number $T$
varied between $1,000$ and $100,000$. The length of the convolution ($L+1$) changed between $2$ and $6$. Comparisons
with the TCC method are shown in Figs.~\ref{fig:LPA-r-3D-geom-celebs}(a)-(d). LPA estimation errors are given in
Table~\ref{tab:LPA-r-3D-geom-celebs}. Figures~\ref{fig:LPA-demo-3D-geom}(a)-(d) and (i)-(l) illustrate the estimations
of the LPA technique on the \emph{3D-geom} and on the \emph{celebrities} databases, respectively.

\begin{table}
    \centering
    \begin{tabular}{|c||@{\hspace{0.1cm}}c|c|c|c|c|}
    \hline
        &$L=1$ & $L=2$ & $L=3$ & $L=4$ & $L=5$\\
    \hline\hline
        3D-geom&$0.20\%(\pm 0.01)$ & $0.20\%(\pm 0.02)$ & $0.19\%(\pm 0.02)$ & $0.20\%(\pm 0.02)$ & $0.20\%(\pm 0.01)$\\
    \hline\hline
        celebrities&$0.33\%(\pm 0.02)$ & $0.33\%(\pm 0.02)$ & $0.34\%(\pm 0.02)$ & $0.34\%(\pm 0.02)$ & $0.34\%(\pm 0.02)$\\
    \hline
    \end{tabular}
    \caption{The Amari-index of the LPA method for database \emph{3D-geom} and $celebrities$, for different convolution lengths: average $\pm$ deviation.
    Number of samples: $T=100,000$. For other sample numbers between $1,000\le T < 100,000$, see Figs.~\ref{fig:LPA-r-3D-geom-celebs}(a) and (c).}
    \label{tab:LPA-r-3D-geom-celebs}
\end{table}

Figures~\ref{fig:LPA-r-3D-geom-celebs}(a) and (c) demonstrate that
the LPA algorithm is able to uncover the hidden components with
high precisions. The Amari-index $r$  decreases according to power
law $r(T)\propto T^{-c}$ $(c>0)$ for sample numbers $T>2000$. The
power law is manifested by straight lines on $\log\log$ scales.
According to Figs.~\ref{fig:LPA-r-3D-geom-celebs}(b) and (d), the
LPA method is superior to the TCC method (i) for all sample
numbers $1,000\le T\le 100,000$, moreover (ii) LPA can provide
reasonable estimates for much smaller sample numbers. This
behavior is manifested by the initial steady increase of the
quotients of the Amari indices of the TCC and LPA methods as a
function of sample number followed by a sudden drop when the
sample number enables reasonable TCC estimations, too. The LPA
method resulted in $1.1-88$-times increase of precision for the
\emph{3D-geom} database and a similar $1.0-87$-times increase for
the \emph{celebrities} database. According to
Table~\ref{tab:LPA-r-3D-geom-celebs}, the Amari-index for sample
number $T=100,000$ is $0.19-0.20\%$ ($0.33-0.34\%$) with small
$0.01-0.02$ ($0.02$) standard deviations for the \emph{3D-geom}
(\emph{celebrities}) database.
\mbox{Figures~\ref{fig:LPA-demo-3D-geom}(e)-(h) and (m)-(q)}
demonstrate that the LPA method may provide acceptable estimations
for reasonably small ($T=20,000$) sample numbers up to convolution
depth $L=20$.

In our test on `\emph{letters}' and `\emph{Beatles}' the number of components and their dimensions were minimal ($d=2$,
$M=2$). According to Figs.~\ref{fig:LPA-r-3D-geom-celebs}(e) and (g), the LPA method found the hidden components. For
the \emph{letters} dataset, the `power law' decline of the Amari-index, that was apparent in the \emph{3D-geom} and the
\emph{celebrities} databases, appears too. For this dataset, Fig.~\ref{fig:LPA-r-3D-geom-celebs}(f) shows that the LPA
method is more precise than the TCC method for all sample numbers. The quotient is between $1.2-110$, and the form of
the curve is similar to those of the \emph{3D-geom} and \emph{celebrities} databases. According to
Table~\ref{tab:LPA-r-ABC}, for sample number $T=75,000$ the Amari-index stays below $1\%$ on average ($0.3-0.36\%$) and
has $0.11-0.15$ standard deviation. Visual inspection of Fig.~\ref{fig:LPA-r-3D-geom-celebs}(g) shows that the LPA
method found the hidden components for sample number $T\ge 30,000$ on the \emph{Beatles} database. We found that the
TCC method gave reliable solutions for sample number $T=50,000$ or so. In addition, according to
Fig.~\ref{fig:LPA-r-3D-geom-celebs}(h) the LPA method is more precise for $T\ge 30,000$ than the TCC technique. The
increase in precision becomes more pronounced for larger convolution parameter $L$. Namely, for sample number $75,000$
and for $L=1,2,5,10,20,30$ the ratios of precision are $1.50, 2.24, 4.33, 4.42, 9.03, 11.13$, respectively on the
average. According to Table~\ref{tab:LPA-r-ABC}, for sample number $T=75,000$ the Amari-index stays below $1\%$ on
average ($0.4-0.71\%$) and has $0.02-0.08$ standard deviation for the \emph{Beatles} test.

\begin{figure}%
\centering%
\subfloat[][]{\includegraphics[height=0.62cm]{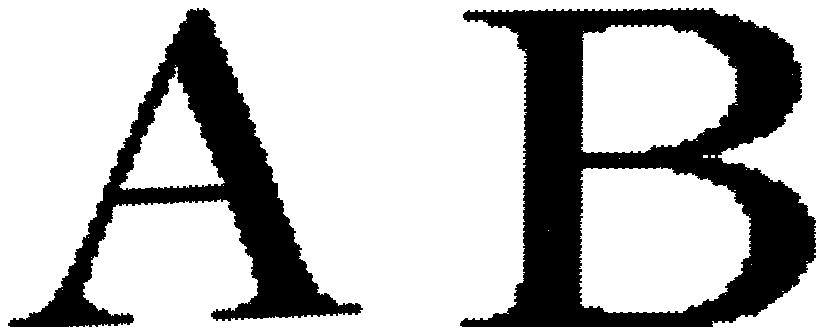}}\hfill%
\subfloat[][]{\includegraphics[height=0.8cm]{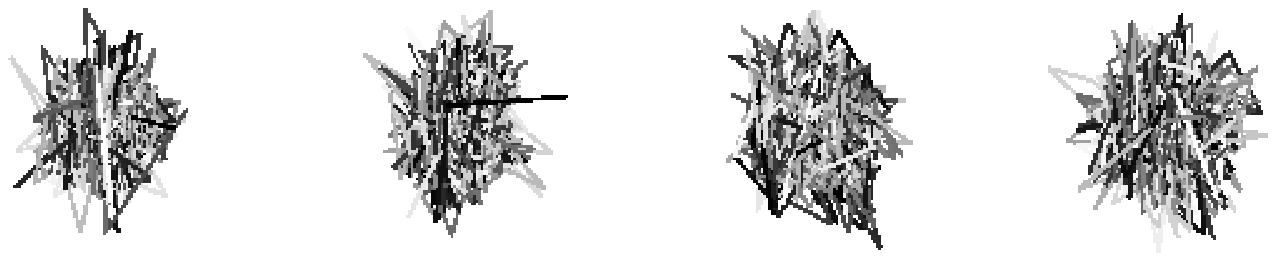}}\hfill%
\subfloat[][]{\includegraphics[height=0.65cm]{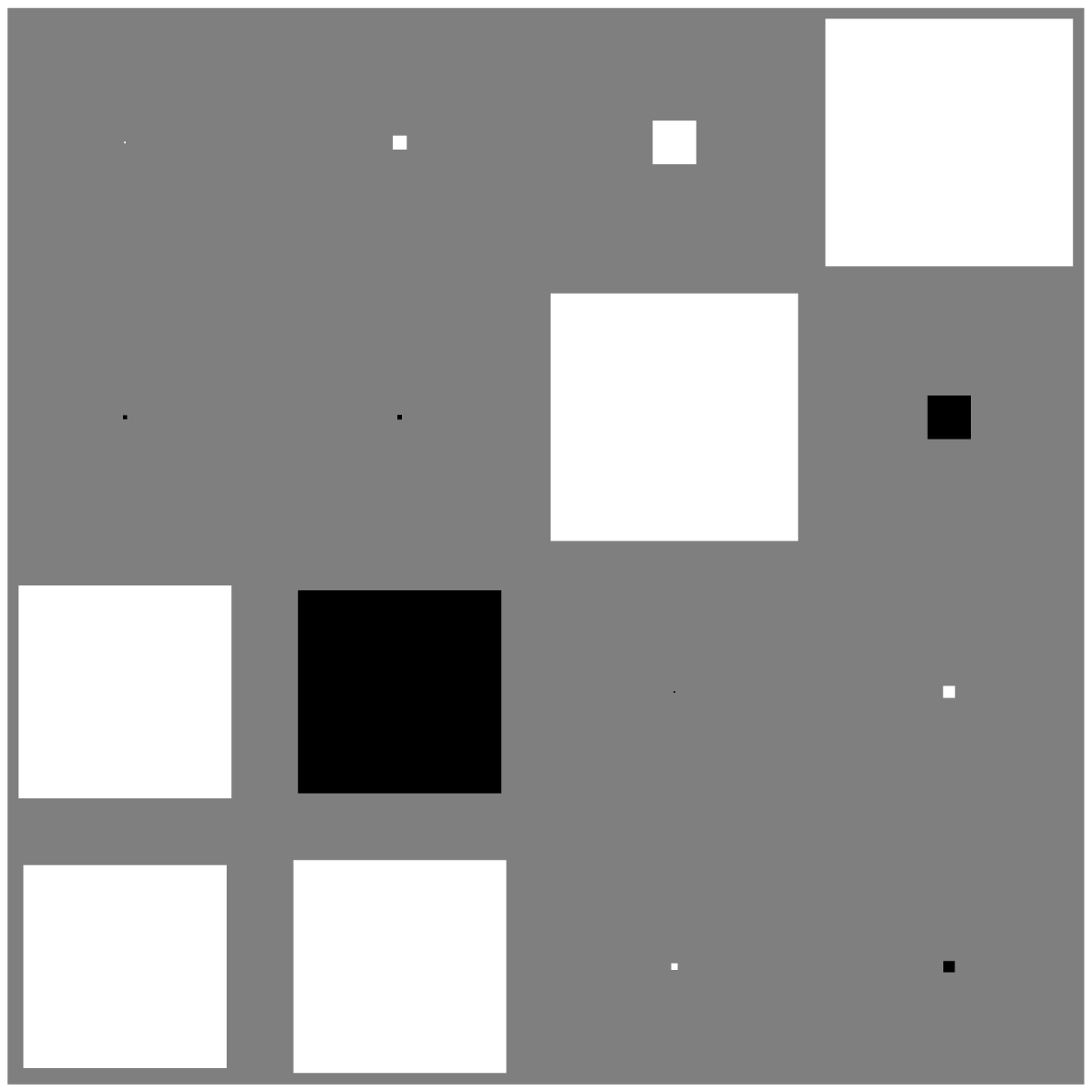}}\hfill%
\subfloat[][]{\includegraphics[height=0.65cm]{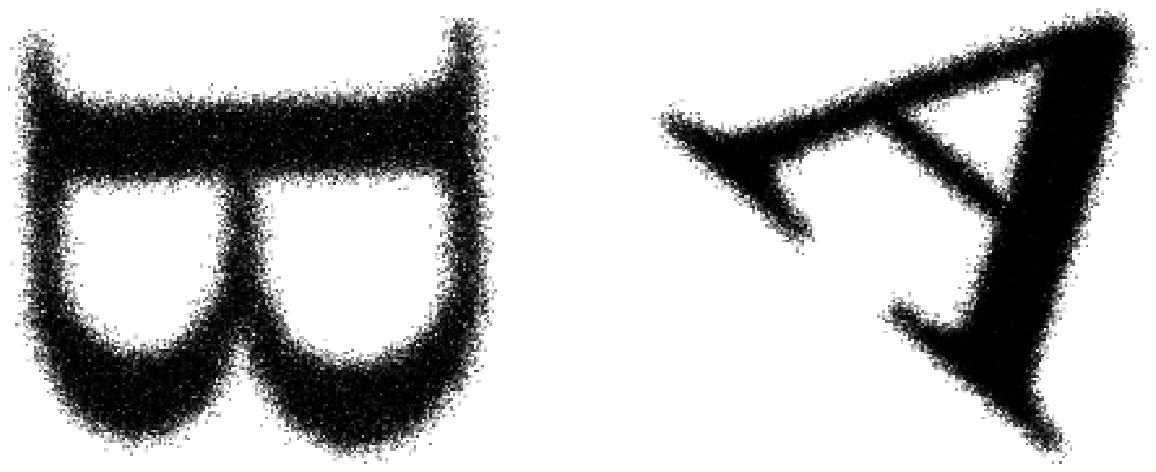}}\\%
\subfloat[][]{\includegraphics[height=1.5cm]{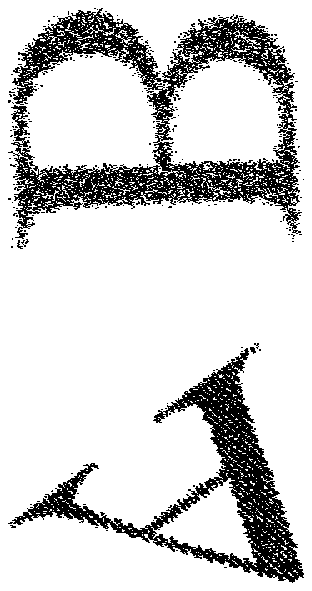}}\hfill%
\subfloat[][]{\includegraphics[height=1.5cm]{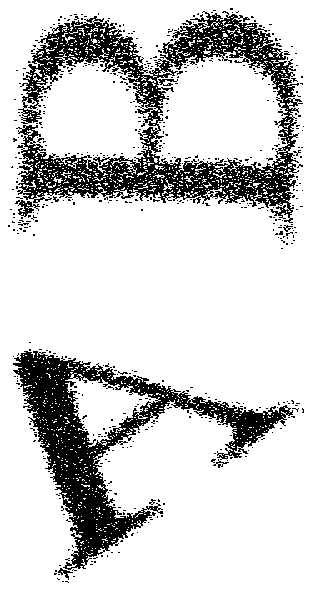}}\hfill%
\subfloat[][]{\includegraphics[height=1.5cm]{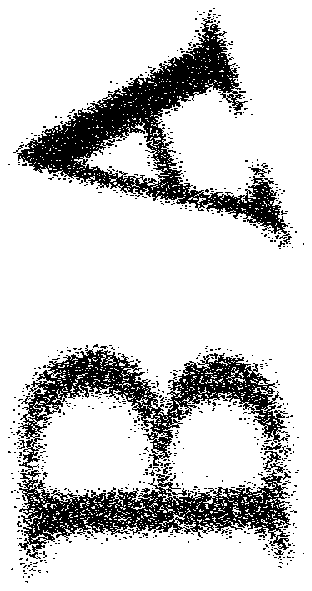}}\hfill%
\subfloat[][]{\includegraphics[height=1.5cm]{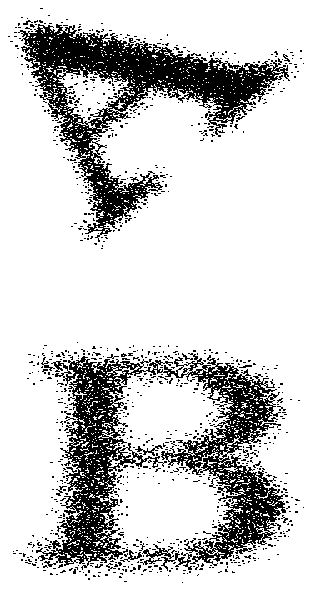}}\hfill%
\subfloat[][]{\includegraphics[height=1.5cm]{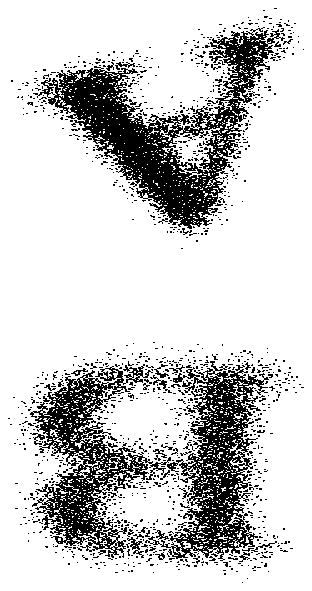}}\hfill%
\subfloat[][]{\includegraphics[height=1.5cm]{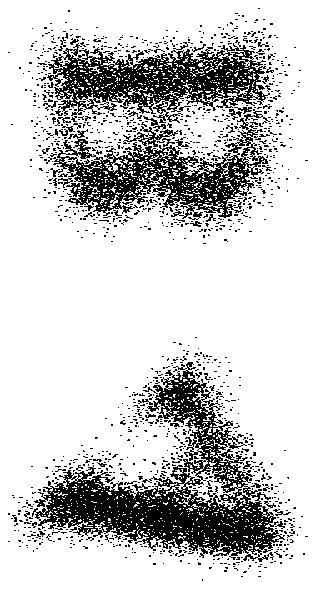}}\hfill%
\subfloat[][]{\includegraphics[height=1.5cm]{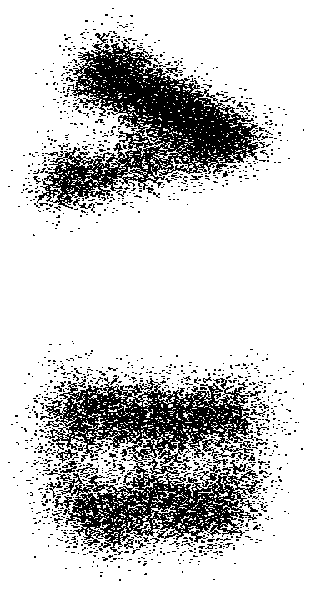}}\hfill%
\subfloat[][]{\includegraphics[height=1.5cm]{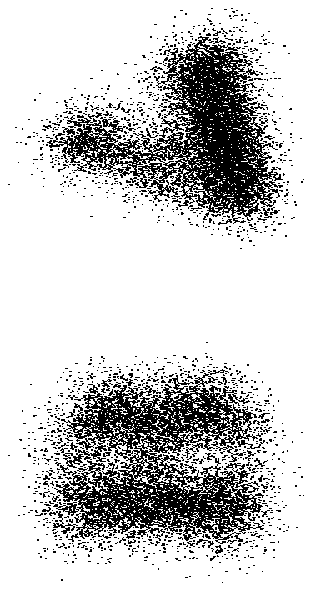}}
\caption[]{Illustration of the LPA method on the uBSSD task for
the \emph{letters} database. \mbox{(a)-(d)}: sample number
$T=75,000$, convolution length $L+1=31$, Amari-index $0.3\%$. (a):
hidden components $\b{s}^m(t)$. (b): observed convolved signals
$\b{x}(t)$, only $1000$ time steps are shown. (c): Hinton-diagram
of $\b{G}$, ideally block-permutation matrix with $2\times 2$
blocks. (d): estimated components ($\hat{\b{s}}^m$). (e)-(l):
dependence of estimated components ($\hat{\b{s}}^m$) on the
convolution parameter $L$. $L$ is $1,5,10,20,50,100,200,230$,
respectively.
Sample number is \mbox{$T=15,000$}.}%
\label{fig:LPA-demo-ABC}%
\end{figure}

\begin{table}
    \centering
    \begin{tabular}{|c|c|c|c|c|c|}
    \hline
        $L=1$ & $L=2$ & $L=5$ &$L=10$ & $L=20$ & $L=30$\\
    \hline\hline
        $0.32\%(\pm 0.11)$ & $0.36\%(\pm 0.14)$ & $0.34\%(\pm 0.13)$ &$0.34\%(\pm 0.15)$ & $0.34\%(\pm 0.11)$ & $0.30\%(\pm 0.14)$\\
    \hline
        $0.71\%(\pm 0.06)$ & $0.64\%(\pm 0.07)$ & $0.53\%(\pm 0.02)$ &$0.75\%(\pm 0.07)$ & $0.45\%(\pm 0.08)$ & $0.40\%(\pm 0.06)$\\
    \hline
    \end{tabular}
    \caption{The Amari-index of the LPA method for database \emph{letters} and \emph{Beatles} for different convolution lengths: average $\pm$ deviation.
    Number of samples: $T=75,000$. First row: \emph{letters}, second row: \emph{Beatles} test.  For other sample numbers between $1,000\le T < 75,000$, see Figs.~\ref{fig:LPA-r-3D-geom-celebs}(e) and (g).}
    \label{tab:LPA-r-ABC}
\end{table}

Both for database \emph{letters} and database \emph{Beatles}, the estimations are acceptable up to about $L=230$
convolution depths for sample number $T=15,000$. We illustrate this in Figs.~\ref{fig:LPA-demo-ABC}(e)-(l) for the
\emph{letters} database with average Amari-index estimations.

\section{Summary}\label{sec:conclusions}
We showed a novel solution method for the undercomplete case of the blind subspace deconvolution (uBSSD) task. We used
a stepwise decomposition principle and reduced the problem with linear prediction to independent subspace analysis
(ISA) task. We illustrated the method on different tests. Our method supersedes the temporal concatenation based uBSSD
method, because (i) it gives rise to a smaller dimensional ISA task, (ii) it produces similar estimation errors at
considerably smaller sample numbers, and (iii) it can treat deeper temporal convolutions.

\bibliographystyle{splncs}

\end{document}